\documentclass{article}
\usepackage[latin1]{inputenc}
\usepackage[T1]{fontenc}

\usepackage{geometry}
\usepackage{graphicx}
\usepackage{upgreek}
\usepackage{isotope}
\usepackage{color}
\usepackage{subfigure}
\usepackage{url}

\newcommand{\spr}{\mbox{$s$-process}}
\newcommand{\sprn}{\mbox{$s$ process}}

\newcommand{\rpr}{\mbox{$r$-process}}

\newcommand{\msun}{\ensuremath{\, M_\odot}}

\newcommand{\beq}{\begin{equation}}
\newcommand{\beqa}{\begin{eqnarray}}
\newcommand{\eeq}{\end{equation}}
\newcommand{\eeqa}{\end{eqnarray}}
\newcommand{\bedis}{\begin{displaymath}}
\newcommand{\edis}{\end{displaymath}}

\begin{document}

\begin{center}

{\huge THE PRODUCTION OF PROTON-RICH}\\
\vspace{0.5cm}
{\huge ISOTOPES BEYOND IRON: }\\
\vspace{0.5cm}
{\huge THE $\gamma$ PROCESS IN STARS}\\
\vspace{1.5cm}

MARCO PIGNATARI \\
\vspace{0.2cm}
{\footnotesize E.A. Milne Centre for Astrophysics, Dept of Physics \& Mathematics, University of Hull,\\ HU6~7RX, United Kingdom.\\
Konkoly Observatory, Research Centre for Astronomy and Earth Sciences, \\Hungarian Academy of Sciences, Konkoly Thege Miklos ut 15-17, H-1121 Budapest, Hungary.\\
NuGrid collaboration, http://www.nugridstars.org\\
mpignatari@gmail.com}\\
\vspace{0.5cm}

KATHRIN G\"OBEL \\
\vspace{0.2cm}
{\footnotesize Goethe University Frankfurt, Max-von-Laue-Str. 1, 60438 Frankfurt am Main, Germany.\\
NuGrid collaboration, http://www.nugridstars.org\\
goebel@physik.uni-frankfurt.de}\\
\vspace{0.5cm}

REN\'E REIFARTH \\
\vspace{0.2cm}
{\footnotesize Goethe University Frankfurt, Max-von-Laue-Str. 1, 60438 Frankfurt am Main, Germany.\\
NuGrid collaboration, http://www.nugridstars.org\\
reifarth@physik.uni-frankfurt.de}\\
\vspace{0.5cm}

CLAUDIA TRAVAGLIO\\
\vspace{0.2cm}
{\footnotesize INAF, Astrophysical Observatory Turin, Strada Osservatorio 20,\\ I-10025 Pino Torinese (Turin), Italy.\\
B$^{2}$FH Association, Turin, Italy.\\
NuGrid collaboration, http://www.nugridstars.org\\
travaglio@oato.inaf.it}

\end{center}

\markboth{Pignatari, G\"obel, Reifarth, Travaglio}{The $\gamma$ process in stars}

\begin{abstract}
Beyond iron, a small fraction of the total abundances in the Solar System is made of proton-rich isotopes, the $p$~nuclei. The clear understanding of their production is a fundamental challenge for nuclear astrophysics. The $p$~nuclei constrain the nucleosynthesis in core-collapse and thermonuclear supernovae. The $\gamma$~process is the most established scenario for the production of the $p$~nuclei, which are produced via different photodisintegration paths starting on heavier nuclei. A large effort from nuclear physics is needed to access the relevant nuclear reaction rates far from the valley of stability. This review describes the production of the heavy proton-rich isotopes by the $\gamma$~process in stars, and explores the state of the art of experimental nuclear physics to provide nuclear data for stellar nucleosynthesis. 
\end{abstract}



\section{Introduction}
\label{intro}

The production of elements heavier than iron in stars is one of the most challenging topics of nuclear astrophysics. The slow neutron capture process ($s$ process, see Ref.~\cite{kaeppeler:11} and references therein) and the rapid neutron capture process ($r$ process, see Ref.~\cite{thielemann:11} and references therein) synthesize the bulk of the heavy elements in the Solar System. The contribution of the intermediate neutron capture process ($i$ process~\cite{cowan:77}) relative to the well established $s$ process and the $r$ process was recently matter of debate, following the identification of $i$-process products in stars at different metallicities~\cite{herwig:11,lugaro:15,dardelet:15,mishenina:15} and in presolar grains~\cite{jadhav:13,fujiya:13,liu:14}.  

Other astrophysical sources from the deepest interior of core-collapse supernovae (CCSNe e.g., Refs.~\cite{woosley:02,langer:12,nomoto:13}) may contribute to the mass region between Fe and Pd. Different neutrino-wind components from the forming neutron star (e.g., Refs.~\cite{hoffman:96,froehlich:06,wanajo:06,farouqi:09,roberts:10,arcones:11}) and the $\alpha$-rich freezout component in CCSN ejecta (e.g., Refs.~\cite{woosley:92,magkotsios:11,pignatari:13a}) may power the production of partially proton-rich or partially neutron-rich isotopes beyond Fe. Their production has been compared to the abundance patterns of old stars born in the early galaxy, which carry the chemical fingerprints of the first generations of stars (e.g., Ref.~\cite{sneden:08}). In particular, they have been proposed to explain the abundance enrichment of Sr, Y and Zr at the neutron shell closure N~=~50 compared to heavier $r$-process elements like Eu (e.g., Refs.~\cite{truran:02,travaglio:04,qian:07,farouqi:10}), and the observed correlation between Ag and Pd in a fraction of stars (e.g., Refs.~\cite{hansen:13}). 
The possible impact of electron-capture SNe~\cite{wanajo:11} and other scenarios like the "cold" $r$ process due to the neutrino spallation on He nuclei in CCSNe~\cite{banerjee:11} need also to be considered. In the future Galactical chemical evolution simulations (see results by Ref.~\cite{hansen:13,ishimaru:06}) should explore consistently the role of all of these processes for the chemical inventory of the galaxy.

\subsection{The classical $p$ nuclei}

Beyond the Fe-group elements, 35 stable proton-rich nuclei\footnote{\ \isotope[74]{Se}, \isotope[78]{Kr}, \isotope[84]{Sr}, \isotope[92,94]{Mo},\isotope[96,98]{Ru}, \isotope[102]{Pd}, \isotope[106,108]{Cd}, \isotope[112,114,115]{Sn}, \isotope[113]{In}, \isotope[120]{Te}, \isotope[124,126]{Xe}, \isotope[130,132]{Ba}, \isotope[136,138]{Ce}, \isotope[138]{La}, \isotope[144]{Sm}, \isotope[152]{Gd}, \isotope[156,158]{Dy}, \isotope[162,164]{Er}, \isotope[168]{Yb}, \isotope[174]{Hf}, \isotope[180]{Ta}, \isotope[180]{W}, \isotope[184]{Os}, \isotope[190]{Pt}, and \isotope[196]{Hg}.} are identified as the $p$~nuclei~\cite{rauscher:13}. Ref.~\cite{cameron:57} discussed their origin and called them {\it excluded isotopes}: they are bypassed by the $s$-process and $r$-process neutron capture paths. The $p$~nuclei are typically 10$-$1000 times less abundant than the more neutron-rich isotopes, and their relative abundance is less than 2\% of the respective element. The exceptions to this rule are the $p$~nuclei \isotope[92]{Mo} and \isotope[94]{Mo} (forming 14.77\% and 9.23\% of the total abundance of Mo in the Sun), and \isotope[94]{Ru} (5.54\% of the Ru in the Sun). However, this does not imply that these isotopes are more abundant in nature than other $p$~nuclei. The most abundant $p$~nuclei in the Solar System is \isotope[74]{Se} (0.89\% of the Se abundance in the Sun). In general, the solar abundances of the light $p$~nuclei between \isotope[74]{Se} and \isotope[96]{Ru} are comparable within a factor of three to four (e.g., Ref.~\cite{lodders:09}). The large differences in the relative concentrations between \isotope[74]{Se}, \isotope[92]{Mo} and \isotope[94]{Mo} result from different nucleosynthesis histories of Se and Mo, which lead to a higher concentration of Se compared to Mo in the Sun.

\subsection{The origin of the $p$ nuclei}

The origin of the $p$~nuclei was investigated starting with the pioneering work of Refs.~\cite{cameron:57,burbidge:57}, and later by Ref.~\cite{audouze:75} and Ref.~\cite{arnould:76}. The $p$~process was originally proposed to be driven by proton captures in the external ejecta of CCSNe~\cite{burbidge:57}. However, first stellar nucleosynthesis simulations for advanced evolution stages of massive stars showed that the production of $p$~nuclei was more likely given by successive neutron-, proton- and $\alpha$-dissociation reactions, with heavier nuclei acting as seeds for lighter ones (e.g., Ref.~\cite{arnould:76}). Ref.~\cite{woosley:78} showed that CCSN explosions in massive stars may naturally provide the conditions to efficiently produce $p$~nuclei via photodisintegration reactions, and called this process $\gamma$~process. We know today that the $\gamma$~process can be activated in both CCSNe and thermonuclear supernovae (SNe~Ia, Ref.~\cite{hillebrandt:13} and references therein). Both scenarios will be discussed in detail in this review.

The $p$~nuclei can be disentangled from neutron-rich isotopes of the same element only in Solar System material due to their low relative concentration. The aim is to reproduce the observed abundances by one or more astrophysical scenarios and to determine their contribution to the galactical chemical evolution. Proton-rich neutrino-wind components like the $\nu$p process are unable to reproduce the $^{92,94}$Mo solar isotopic ratio~\cite{fisker:09,bliss:15}. Therefore, they do not play a dominant role in the production of the solar inventory of these species. 
A powerful nucleosynthesis process working in the same mass region is the $rp$ process, which is activated in neutron stars accreting H-rich and He-rich material. This process is recurrently activated at the surface of the compact object during H-burning runaway events, leading to the production of X-ray bursts~\cite{wallace:81,schatz:98,schatz:99,galloway:08,thielemann:10}. The $rp$ process is characterized by a sequence of proton captures close to the proton dripline and $\beta$-decays. Its nucleosynthesis path terminates at the proton magic Sn isotopes (Z~=~50) and at the SnSbTe cycle, where the reaction $^{107}$Te($\gamma$,$\alpha$)$^{103}$Sn guides the flow back to Sn~\cite{schatz:01}. Therefore, the $rp$ process may lead to huge overproductions of $p$~nuclei up to Cd, although it typically stops at lighter elements in stellar simulations (e.g., Refs.~\cite{woosley:04,heger:07,fisker:08,parikh:13}). However, according to stellar simulations, the $rp$-process products are not ejected, and they do not contribute to the abundances in the Solar System (e.g., Ref.~\cite{fisker:08}). We will neglect these processes in the further discussion because of the negligible or uncertain contributions to the solar abundances, and focus on the $\gamma$~process.

\subsection{The production of more neutron-rich isotopes by the $\gamma$ process}

The processes responsible for the production of $p$~nuclei may also contribute to the production of more neutron-rich isotopes. Ref.~\cite{kaeppeler:82} provided a phenomenological evaluation of this contribution by comparing the solar abundances of $p$~nuclei with neighboring species belonging to the same element. According to this simple recipe, e.g., about 20\% of the solar $^{128}$Xe is of $\gamma$-process origin, while this isotope was originally considered as an $s$-only isotope. Observations of Xe in samples of presolar silicon carbide grains of mainstream type later confirmed these findings~\cite{zinner:14}. These grains condensed in the envelope of old Asymptotic Giant Branch stars, and still carry a pure $s$-process component from their parent stars (see e.g., Refs.~\cite{lewis:90,gallino:90,lewis:94}). For Xe, the $s$-process production of $^{128}$Xe was indeed lower than the $s$-process component of the other Xe $s$-process isotope $^{130}$Xe~\cite{reifarth:04,pignatari:06}. The $s$-process isotope $^{80}$Kr is another example, which may have an important explosive contribution in \mbox{CCSNe}~\cite{tur:09}. Empirical laws to reproduce the solar abundance distribution of $p$~nuclei and their abundances relative to the $s$-process nuclei were discussed by, e.g., Ref.~\cite{hayakawa:08}.

\subsection{$s$- and $r$-process contributions to the $p$ nuclei}

Originally, the 35 $p$~nuclei were identified assuming that they are bypassed by the $s$~and the $r$~processes. Today is quite well accepted that the $p$~nuclei \isotope[152]{Gd} and \isotope[164]{Er} receive a dominant \spr\ contribution from low-mass AGB stars~\cite{arlandini:99,bisterzo:11}. In particular, the $s$-process production of $^{164}$Er is driven by the $\beta$-decay channel of $^{163}$Dy, which becomes unstable at stellar temperatures~\cite{takahashi:87}. The bulk of the isotopes \isotope[113]{In} and \isotope[115]{Sn} is not produced by the $\gamma$~process. Nuclear uncertainty studies for the $\gamma$~process in CCSNe did not solve this puzzle~\cite{rauscher:06,rapp:06}. Their origin today is still unclear. A potential contribution from the $r$~process has been proposed, but more work is needed to solve this mystery~\cite{nemeth:94,dillmann:08}.
Finally, the $\gamma$~process can only partially reproduce the abundances of \isotope[138]{La} and \isotope[180]{Ta}. Indeed, \isotope[138]{La} includes a significant contribution from neutrino-capture reactions on \isotope[138]{Ba}~\cite{woosley:90,goriely:01}. The long-lived \isotope[180]{Ta} isomer (half-life larger than $1.2\times10^{15}$ yr, see Ref.~\cite{cumming:85}) may be efficiently produced by neutrino-spallation reactions on \isotope[180]{Hf}~\cite{byelikov:07,sieverding:15} and by the \sprn\ in low-mass AGB stars (see Refs.~\cite{arlandini:99,goriely:00,bisterzo:11} for different and controversial predictions). \isotope[180]{Ta} is also fed by the branching at $^{179}$Hf, which is a stable isotope that becomes unstable at stellar temperatures, opening a small branching~\cite{mohr:07}.

\subsection{About this review}

The $p$~nuclei observed in the Solar System have been produced by more than one process. The astrophysical scenarios are especially uncertain for the light $p$~nuclei, where many processes may contribute to their abundances. We refer to Refs.~\cite{meyer:94,arnould:03,rauscher:13} for previous and exhaustive reviews of the different kinds of $p$ processes in stars. 

This review aims at presenting the main features of the $\gamma$~process in stars. The comparison of theoretical stellar simulations with available observations provides the basis to constrain stellar models and nuclear data. We summarize the observations of $p$~nuclei in the Solar System in Section~\ref{sec:observations}. We explore the $\gamma$~process in its two main astrophysical sites, core-collapse supernovae CCSNe in Section~\ref{sec:pprocess_massive} and thermonuclear supernovae SNe~Ia in Section~\ref{pprocess-snia}. Stellar $\gamma$-process simulations need reliable nuclear data in order to obtain robust abundance yields. This may be extremely challenging, since the $\gamma$-process nucleosynthesis path moves far from the valley of stability. The involved reaction rates have only become accessible to experimental measurements in the last years, or are still beyond the capability of present technologies. The nuclear physics involved in the $\gamma$~process will be discussed in Section~\ref{nuclear: rene}. Final discussions and conclusions are given in Section~\ref{sec: conclusions}.

\section{Observations from the Solar System}
\label{sec:observations}

The solar isotopic distribution of the $p$~nuclei~\cite{lodders:09} is the key observational constraint for the $\gamma$ process and all the other $p$ processes. The distribution is composed of solar spectroscopic data and investigations of meteoritic compositions. It is the fundamental benchmark for stellar simulations to test our understanding of the production of $p$~nuclei in stars. 

There is only limited information about the isotopic abundances from the spectroscopic data of other stars than the Sun. Elemental abundances are available for a few exceptions: Abundances of molybdenum can be observed in old metal-poor stars~\cite{peterson:13}. However, the abundances could result from the $p$~nuclei \isotope[92]{Mo} and \isotope[94]{Mo} or from the $r$-process isotope \isotope[100]{Mo}. We cannot disentangle the abundances of the different isotopes, but only guess that the star has the same Mo isotopic distribution as the Sun.

\subsection{Signatures of extinct radionuclides in meteorites}

Extinct radionuclides found in meteorites are an important observable to investigate $p$-process nucleosynthesis~\cite{dauphas:11,davis:14,lugaro:16}. Radionuclides like \isotope[92]{Nb} and \isotope[146]{Sm} are formed on the nuclear reaction path in the $p$ process. Their signatures were detected as an excess abundance of the daughter nuclei \isotope[92]{Zr} and \isotope[146]{Nd}. 

The isotope \isotope[92]{Nb} is produced by the $\gamma$ process, but is completely shielded from contributions from $rp$ or ${\nu}p$ processes~\cite{dauphas:03}. As such, it can help test models of $p$-process nucleosynthesis. Meteorite measurements show that this nuclide was present at the birth of the Solar System (with an initial \isotope[92]{Nb}/\isotope[92]{Mo} ratio of (2.80$\pm$0.5)$\times$10$^{-5}$)~\cite{harper:96,schonbachler:02,rauscher:13}. Its astrophysical production site is still unknown. Recently, Ref.~\cite{travaglio:14} found that \isotope[92]{Nb} could be produced in SNIa.

The production of \isotope[146]{Sm} in nucleosynthesis calculations is influenced by uncertainties in its half-life and the involved nuclear physics. Ref.~\cite{travaglio:14} calculated a \isotope[146]{Sm}/\isotope[144]{Sm} ratio for SNIa which is compatible with the meteoritic value when using a $^{148}$Gd($\gamma$,$\alpha$)$^{144}$Sm rate based on a recent recalculated ($\alpha$,$\gamma$) cross section by Ref.~\cite{rauscher:13}, and Ref.~\cite{rauscher:13} concluded that the recalculated rate does not help to fit the meteoritic value for CCSNe. Hence, SNe~Ia seem to be the favored production sites for \isotope[146]{Sm}. However, this scenario is hard to reconcile with the signature of another extinct radionuclide, \isotope[53]{Mn}~\cite{lugaro:16}. 

The isotopes \isotope[97,98]{Tc} have not been detected in meteorites so far. Since only upper limits have been derived, it is not helpful yet to address $p$-nucleosynthesis models.

\subsection{Stellar dust}

Nucleosynthesis signatures involving $p$~nuclei can also be identified in stellar dust. The dust had been made by other stars before the Sun was formed. Primitive meteorites carry several types of dust of presolar origin, coming from different stellar sources~\cite{zinner:14}. Part of them condensed in CCSNe: low-density graphite grains, silicon carbide grains of type X~\cite{besmehn:03} and type C~\cite{hoppe:12,pignatari:13b}, maybe some classified as nova and type AB~\cite{pignatari:15}, as well as nano-diamonds.

Presolar nano-diamonds are the most abundant type of presolar grains ($\sim$1000 part per milion)~\cite{huss:95}. They typically have the size of a few nanometers, and consist of about $10^3$ C atoms. Other trace elements have a concentration of about one atom per million grains~\cite{koscheev:01}. Hence, abundance measurements of the trace elements require to analyze a large sample of grains~\cite{huss:95}. 

The presolar nano-diamonds carry the Xe-HL component~\cite{lewis:87}, which was first identified in the Allende meteorite by Ref.~\cite{lewis:75}. Compared to solar concentrations, the Xe-HL signature is made by enhanced {\it light} and {\it heavy} stable nuclei: \isotope[124,126]{Xe} (Xe-L) and \isotope[134,136]{Xe} (Xe-H). 
Since the \isotope[124,126]{Xe} present in the Solar System were made by the $\gamma$ process, and \isotope[134,136]{Xe} are considered as \rpr\ isotopes~\cite{thielemann:11}, it was established that the presolar diamonds carrying the Xe-HL condensed in CCSNe ejecta. 
Xe-L cannot be disentangled from Xe-H since the diamonds carrying the two components are well mixed. The corresponding process cannot be explained so far. Furthermore, diamonds are carbon-rich grains, while the $\gamma$ process is activated in oxygen-rich stellar layers, where carbon-rich dust should not form (however, see discussion in Ref.~\cite{clayton:99,ebel:01}). Last but most importantly, the isotopic ratio of the Xe-L isotopes is not consistent with the same ratio of these $p$~nuclei in the solar system~\cite{amari:09,ott:12}. It is unknown why they are different.

\subsection{Anomalies in meteorites}

Isotopic anomalies have been found in different types of meteoritic material, where different types of presolar dust could have been the pristine carriers~\cite{dauphas:04}. Anomalies have been confirmed for the $p$~nuclei $^{92,94}$Mo (e.g., Refs.~\cite{dauphas:02,burkhardt:11}), $^{96,98}$Ru (e.g., Refs.~\cite{chen:10,fischer-godde:15}), $^{144}$Sm (e.g., Refs.~\cite{andreasen:06,qin:11}) and $^{180}$W (Refs.~\cite{schulz:13,peters:14,cook:14}). Compared to the average abundances in the Solar System, anomalies for some $p$~nuclei have also been measured in chondritic meteorites (both in the bulk material and in Ca-Al-rich inclusions) and in old planetesimal fragments like iron meteorites. Clear $^{84}$Sr heterogeneities are reported only in Ca-Al-rich inclusions~\cite{moynier:12,hans:13,paton:13}, as well as for $^{138}$La~\cite{shen:03,chen:15}. Ref.~\cite{mayer:15} also reported potential $^{102}$Pd anomalies in IVB iron meteorites, but errors were too large to derive any significant conclusions. For most of these cases, abundance signatures may be explained by $s$-process deficits in the parent body compared to the solar composition, instead of the contribution from one or more $p$~processes (e.g., Ref.~\cite{dauphas:04}).

At present, $^{144}$Sm is the only $p$~nucleus where the measured anomalies cannot be explained by an $s$-process deficit. The discussion is controversial but still compatible with the scenario for $^{180}$W. The anomalies could be explained by a $\gamma$-process anomaly. The variations are small~\cite{qin:11}, but for sure this is an interesting case for further studies.

Some anomalies for Mo and Ru $p$~nuclei compared to the solar abundances have been measured in single silicon carbide grains of type X. Ref.~\cite{hallmann:13} proposed that such anomalies are compatible with nucleosynthesis signatures from neutrino-driven winds from CCSNe. This scenario might also explain other anomalies in Mo, e.g., the \isotope[95,97]{Mo} excess compared to solar abundances~\cite{hallmann:13,bliss:15}. However, also silicon carbide grains require to condense in C-rich material~\cite{ebel:01}, and isotopic signatures for light elements seem to be compatible with C-rich material from explosive He burning regions~\cite{pignatari:13c,pignatari:15} and not with O-rich material~\cite{lin:10}. It is unclear how material carrying the signature of neutrino winds from the deepest CCSN ejecta might pollute the explosive He-burning region outward in the ejecta without affecting also the isotopic abundances of light and intermediate elements. Furthermore, the \isotope[95,97]{Mo} excesses can be also explained by the neutron-burst in the explosive He shell, triggered by the \isotope[22]{Ne}($\alpha$,n)\isotope[25]{Mg} reaction during the SN shock passage~\cite{meyer:00}.
A more careful analysis is needed to extract information about one or more $p$ processes from the observed anomalies in Mo and Ru $p$~nuclei.
 

\section{The production of $p$ nuclei in massive stars}
\label{sec:pprocess_massive}

\begin{figure}
\centering
\includegraphics[width=0.8\textwidth]{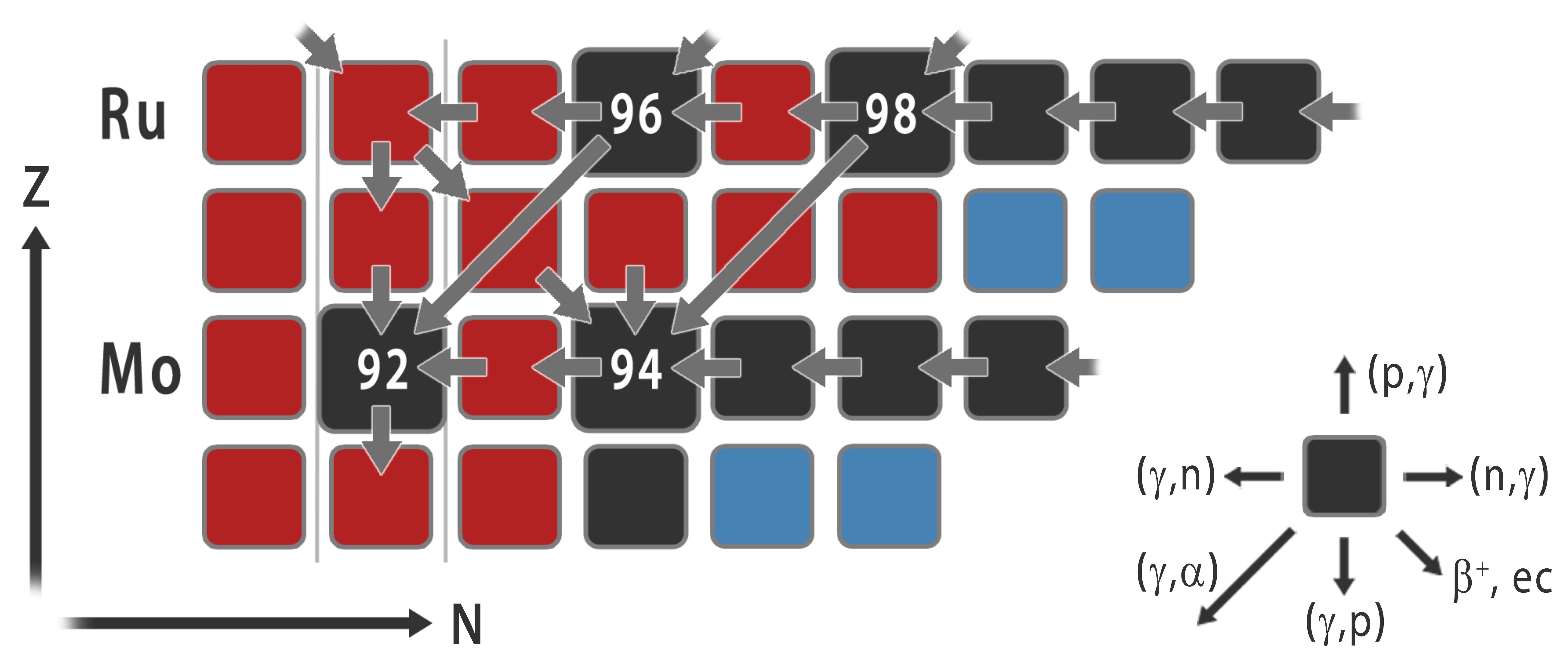}
\caption{Nucleosynthesis flow leading to the production of $p$ nuclei of molybdenum and ruthenium. Neutron-, proton- and $\alpha$-dissociation reactions on more neutron-rich nuclei lead to the proton-rich isotopes during the $\gamma$ process. The p-nuclei are produced directly or via weak reactions ($\beta$-decays and electron captures) on unstable nuclei.}
\label{fig:gProcessReacPaths} 
\end{figure}

The \mbox{$p$ nuclei} are produced in sequences of photodisintegration reactions starting at $r$- and $s$-process nuclei as seed in the $\gamma$ process. The required temperatures range from 2.0 to \mbox{3.5$\times$10$^{9}$ K}. Neutron-, proton- and $\alpha$-dissociation reactions on more neutron-rich nuclei drive the masses towards the proton-rich side of the valley of stability, reaching the \mbox{$p$ nuclei} directly or via weak reactions ($\beta$-decays and electron captures) on unstable nuclei (see Fig.~\ref{fig:gProcessReacPaths}).

\subsection{Astrophysical sites}

Ref.~\cite{arnould:76} first proposed the oxygen burning phase during the pre-supernova evolution of massive stars as a possible site for the $\gamma$ process. Refs.~\cite{thielemann:85,woosley:02} argued that the oxygen-burning stellar layers are further processed by Si burning in the center and are buried in the forming neutron star. The oxygen burning layers are exposed to extreme conditions during the SN shock passage, where all the pre-supernova $p$~nuclei are destroyed (e.g., Refs.~\cite{rauscher:02,pignatari:13}). However, Ref.~\cite{rauscher:02} stated that this is not always the case. Depending on the complex evolution of the stellar structure in the advanced evolution stages, the $p$ nuclei produced in O-burning layers might be mixed into outer layers and ejected by the SN explosion together with the explosive $\gamma$-process component. Therefore, the effective impact of the pre-explosive $\gamma$-process yields depends on the explosion mechanism and on the stellar structure behavior in the last days before core collapse. The occurrence of extensive mixing between different burning shells (in particular, mixing between convective O shells and convective C shells, e.g., Refs.~\cite{rauscher:02,arnett:11}) may significantly enhance the final $\gamma$-process yields.

The $\gamma$ process during a CCSN explosion is the most well-established astrophysical scenario for the nucleosynthesis of the $p$ nuclei~\cite{woosley:78}. Since earlier works~\cite{rayet:90,prantzos:90,rayet:95}, the O/Ne-rich layers of massive stars were considered to host the $\gamma$ process. The $\gamma$ process is activated with typical timescales of less than a second when the shock front passes through the O/Ne burning zone.

\subsection{Production of $p$ nuclei in a 1D CCSN model}

\begin{figure}
\centering \mbox{
\subfigure
{\includegraphics[width=0.5\textwidth]{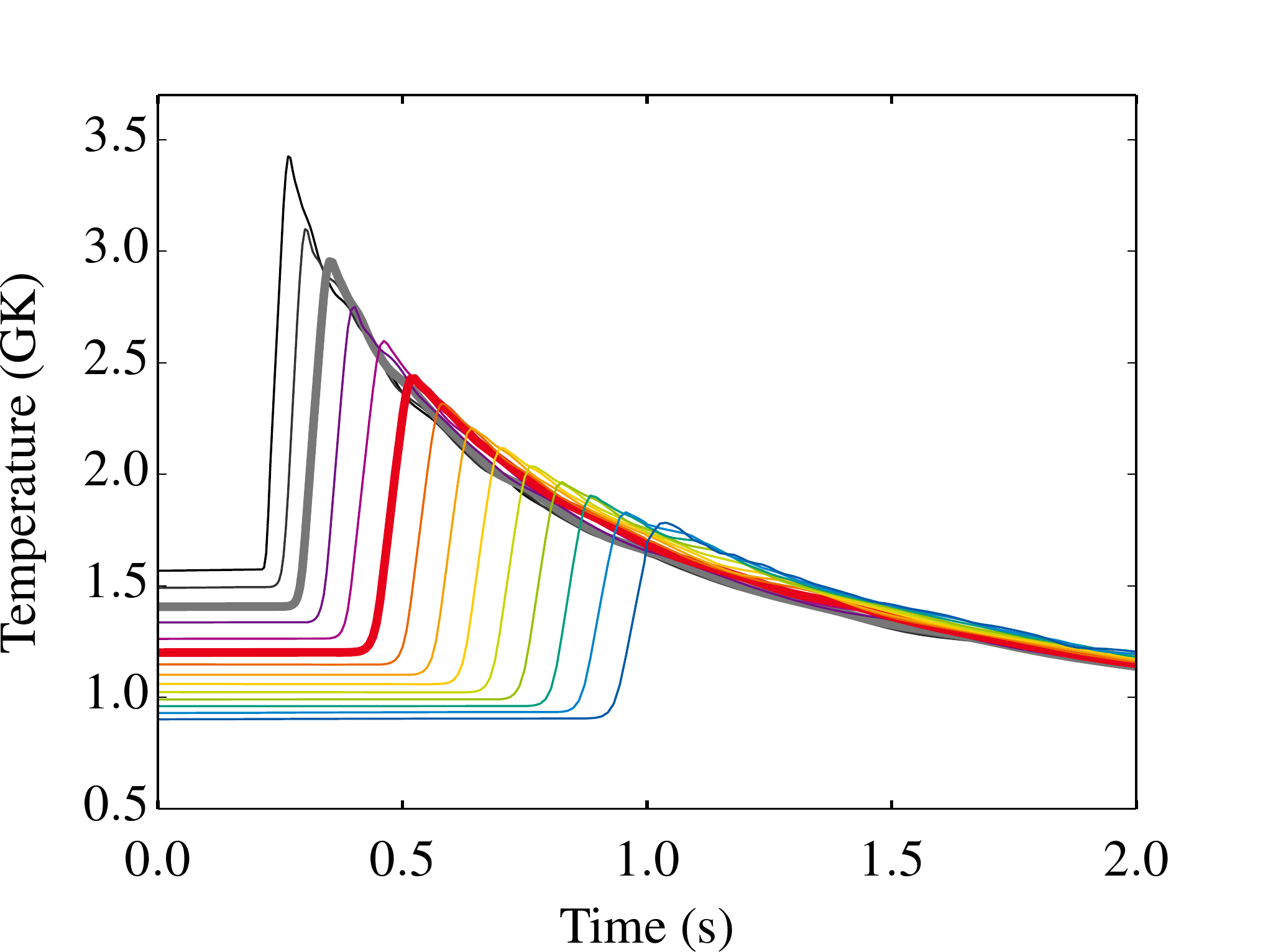}}
{\includegraphics[width=0.5\textwidth]{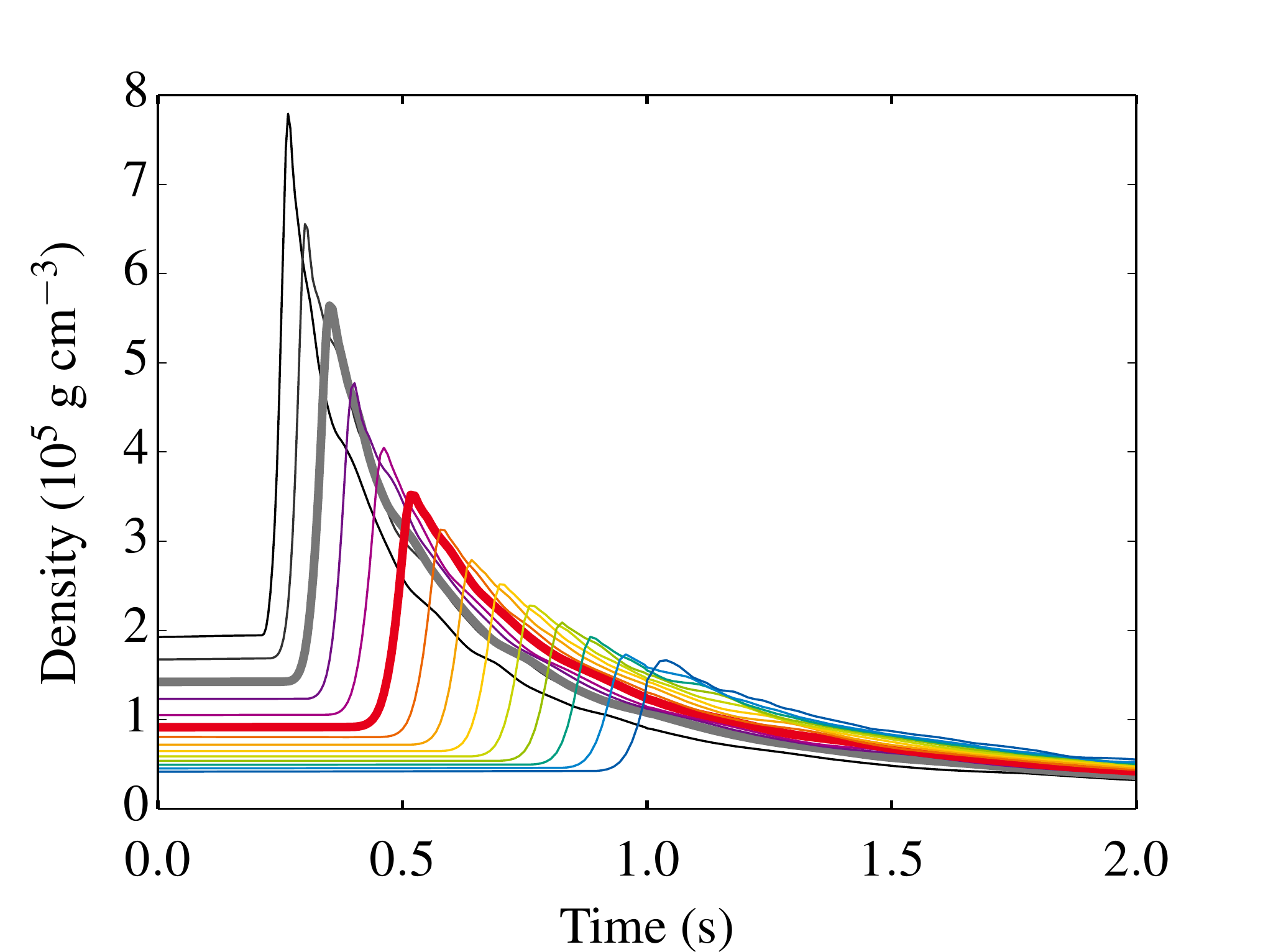}}}
\caption{Temperature and density profiles of the Hashimoto~\cite{rapp:06} trajectories. Left to right: profiles from the innermost mass layer to the outermost one. The shock front reaches the mass layers successively. Trajectories 3 and 6, which will be discussed in further detail, are marked with thick lines.}
\label{fig:HashTraj} 
\end{figure}

We discuss the details of the $\gamma$ process for a one-dimensional CCSN model. In the model used by Ref.~\cite{rapp:06} in their analysis, the shock front passes through the O/Ne burning zone, which is divided in 14 representative mass zones. The innermost mass layer provides the hottest and densest environment with a peak temperature of \mbox{3.45 GK} and a peak density of \mbox{7.85$\times$10$^{5}$g/cm$^3$}. The maximum values drop continuously to 1.79 GK and 1.68$\times$10$^{5}$g/cm$^3$ for the outermost mass layer. Fig.~\ref{fig:HashTraj} shows the temperature and density profiles of the so-called "Hashimoto trajectories", which describe the astrophysical environment of each layer as a function of time. The shock front reaches the mass layers successively. Temperature and density rapidly increase to the maximum values and afterwards drop slowly. 

\begin{figure}
\centering
\includegraphics[width=0.85\textwidth]{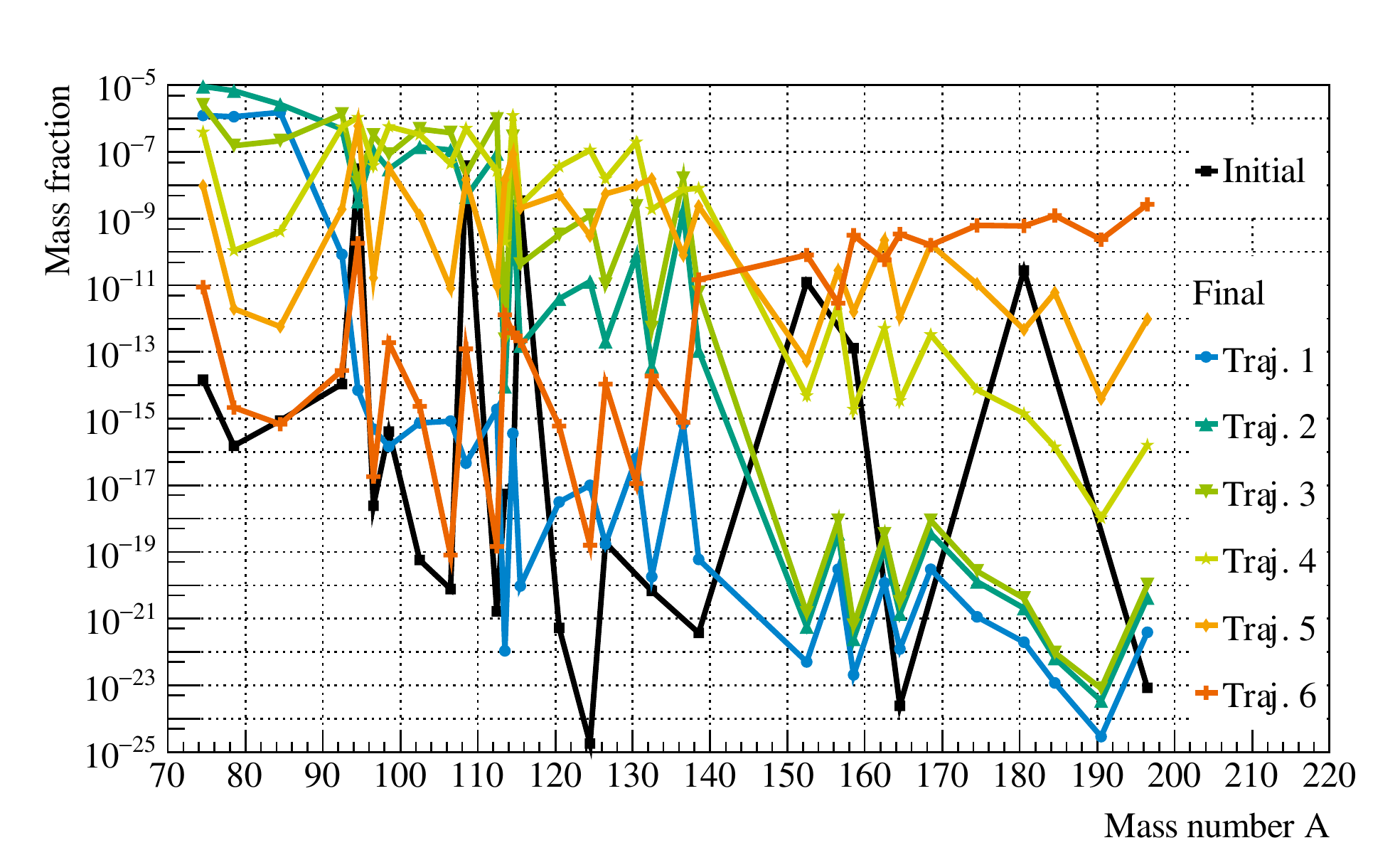}
\caption{Initial and final mass fractions of all \mbox{p-nuclei} for Hashimoto trajectories 1 to 6.}
\label{fig:HashAbu} 
\end{figure}

Fig.~\ref{fig:HashAbu} gives an overview of the production of the $p$ nuclei in the different mass layers, and, hence, for the different temperature profiles of the one-dimensional CCSN model. The heavy $p$ nuclei are destroyed in the hottest environments, and the masses are shifted towards the light $p$ nuclei and even below. The intermediate mass $p$ nuclei \mbox{($A$ $\approx$ 92-136)} are produced significantly in the peak temperature range of about 2.5 to 3.0~GK. The heavy $p$ nuclei \mbox{($A$ $\ge$ 140)} are produced at moderate peak temperatures below 2.5~GK. Beyond these general trends the relative contributions to the total integrated abundances differ for each isotope. 

The abundances, and the absolute and relative nucleosynthesis fluxes give detailed and quantitative information about the production (or destruction) of a single isotope. Absolute nucleosynthesis fluxes show the main reaction paths for a certain time range in the simulation. The reaction rate and the abundance of the parent species determine the nucleosynthesis flux~\cite{herwig:08a}. Relative fluxes for a single isotope represent the relative net abundance yield in the simulation. They also illustrate the most relevant reaction paths for the nucleus~\cite{gobel:15}.

\begin{figure}
\centering
\resizebox{10cm}{!}{\rotatebox{0}{\includegraphics{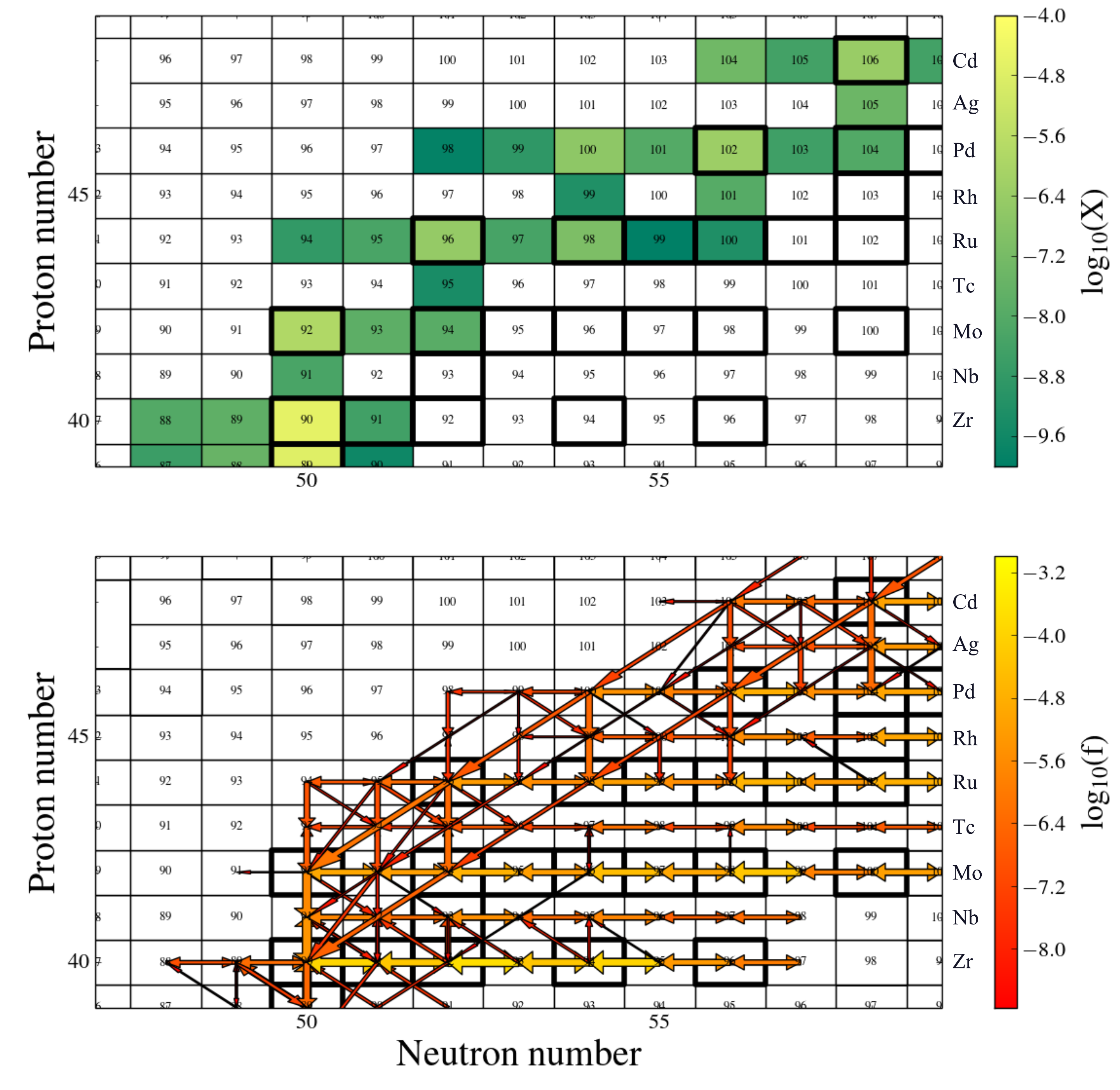}}}
\resizebox{10cm}{!}{\rotatebox{0}{\includegraphics{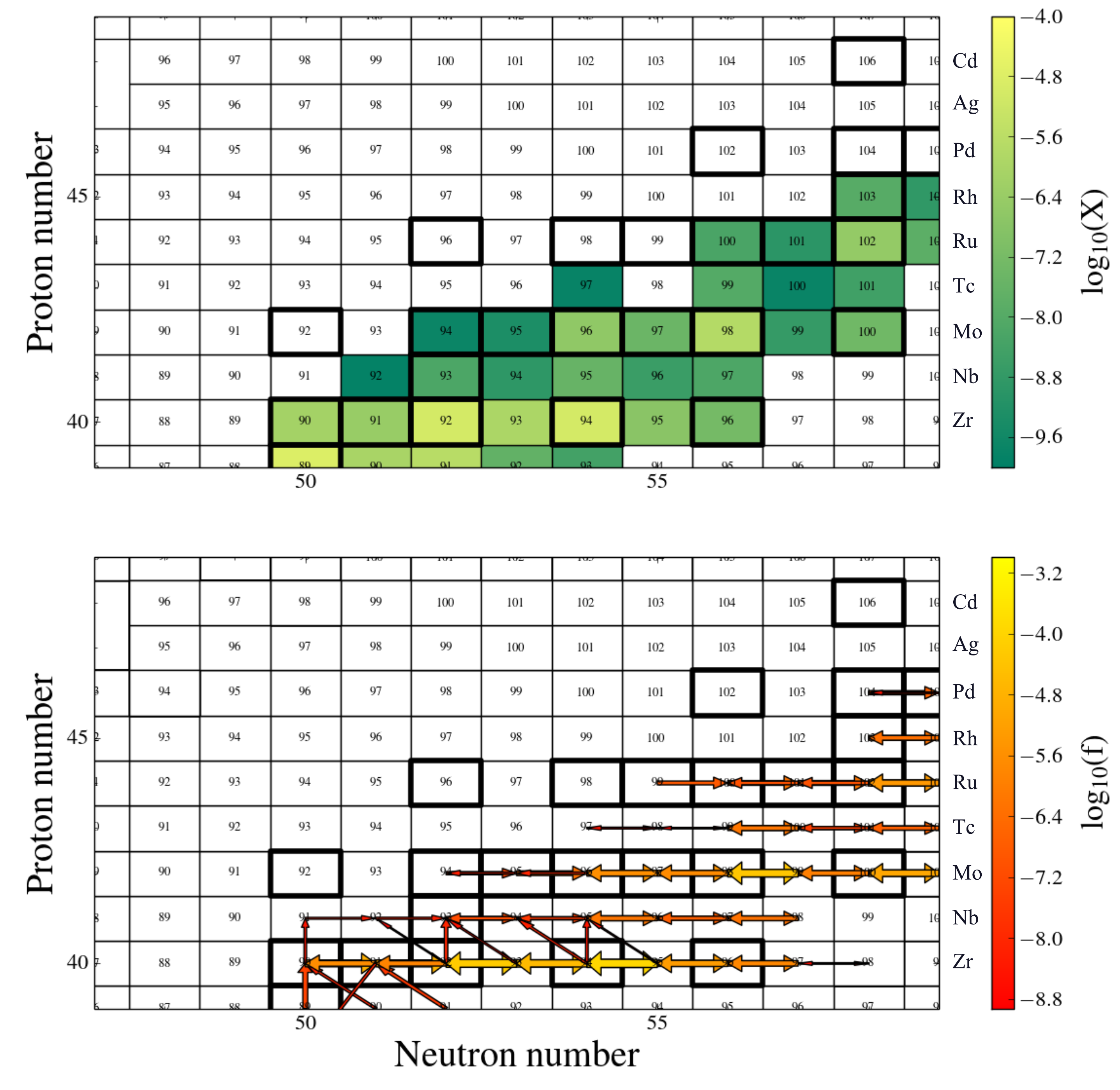}}}
\caption{Mass fraction distributions (green) and time-integrated nucleosynthesis fluxes (arrows with red to yellow color) in the mass region between Zr and Cd for Hashimoto trajectory 3 (upper panels, peak temperature T~=~2.95~GK) and 6 (lower panels, peak temperature T~=~2.43~GK). The abundances are given at the end of the simulation, when some unstable isotopes are still present. The nucleosynthesis fluxes, [$\delta Y_{\rm i}$/$\delta$t]$_{\rm j}$, show the variation of the abundance $Y_{\rm i}$ = $X_{\rm i}$/$A_{\rm i}$ due to the reaction j. The arrow width and color corresponds to the flux strength. Heavy-lined boxes correspond to the stable isotopes.}
\label{fig:abund_and_flux_zr-cd}
\end{figure}

As an example, Fig.~\ref{fig:abund_and_flux_zr-cd} shows the isotopic abundances and the time-integrated absolute nucleosynthesis fluxes for Hashimoto trajectories 3 (upper panel) and 6 (lower panel) in the mass region between Zr and Cd, where the $p$ nuclei $^{92,94}$Mo, $^{96,98}$Ru, $^{102}$Pd and $^{106}$Cd are located. The temperature profiles reach their peaks at 2.95~GK (traj. 3) and 2.43~GK (traj. 6). In the hot environment, the heaviest material is photodisintegrated and is feeding lighter nuclei (compare also Fig.~\ref{fig:HashAbu}), in particular the mass region in the figure. The nucleosynthesis paths show the complexity of this process, where multiple reaction channels are activated. Neutron-dissociation reactions drive the masses to the proton-rich nuclei, while proton- and $\alpha$-dissociation reactions lead to lighter elements. In the cooler environment, the neutron-dissociation and some proton-dissociation reactions are already activated, but they are not strong enough to significantly produce the lightest $p$~nuclei in the isotopic chains.

\begin{figure}
\centering \mbox{
\subfigure
{\includegraphics[trim=150 140 80 250, clip=true,width=0.5\textwidth]{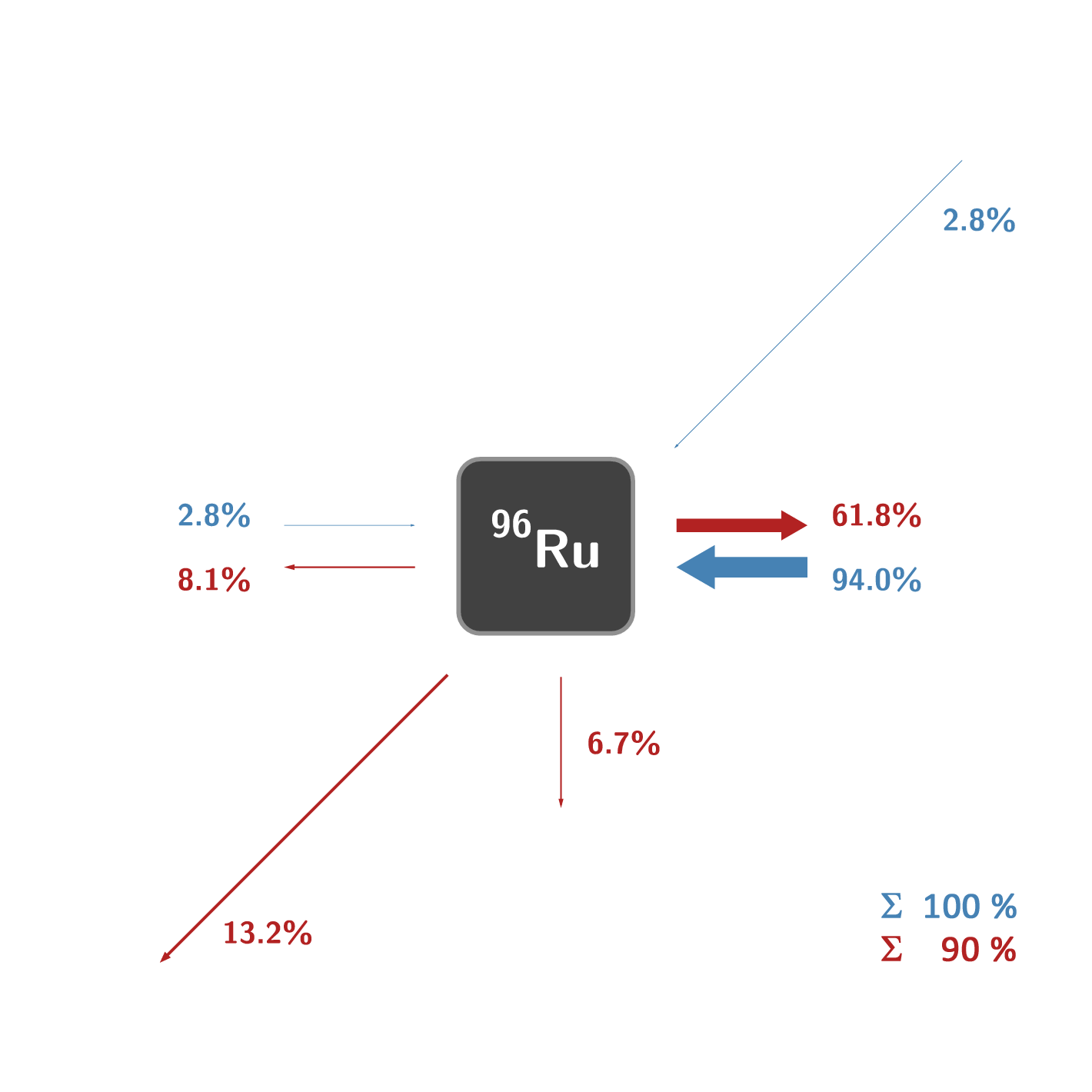}}
\subfigure
{\includegraphics[trim=150 140 80 250, clip=true,width=0.5\textwidth]{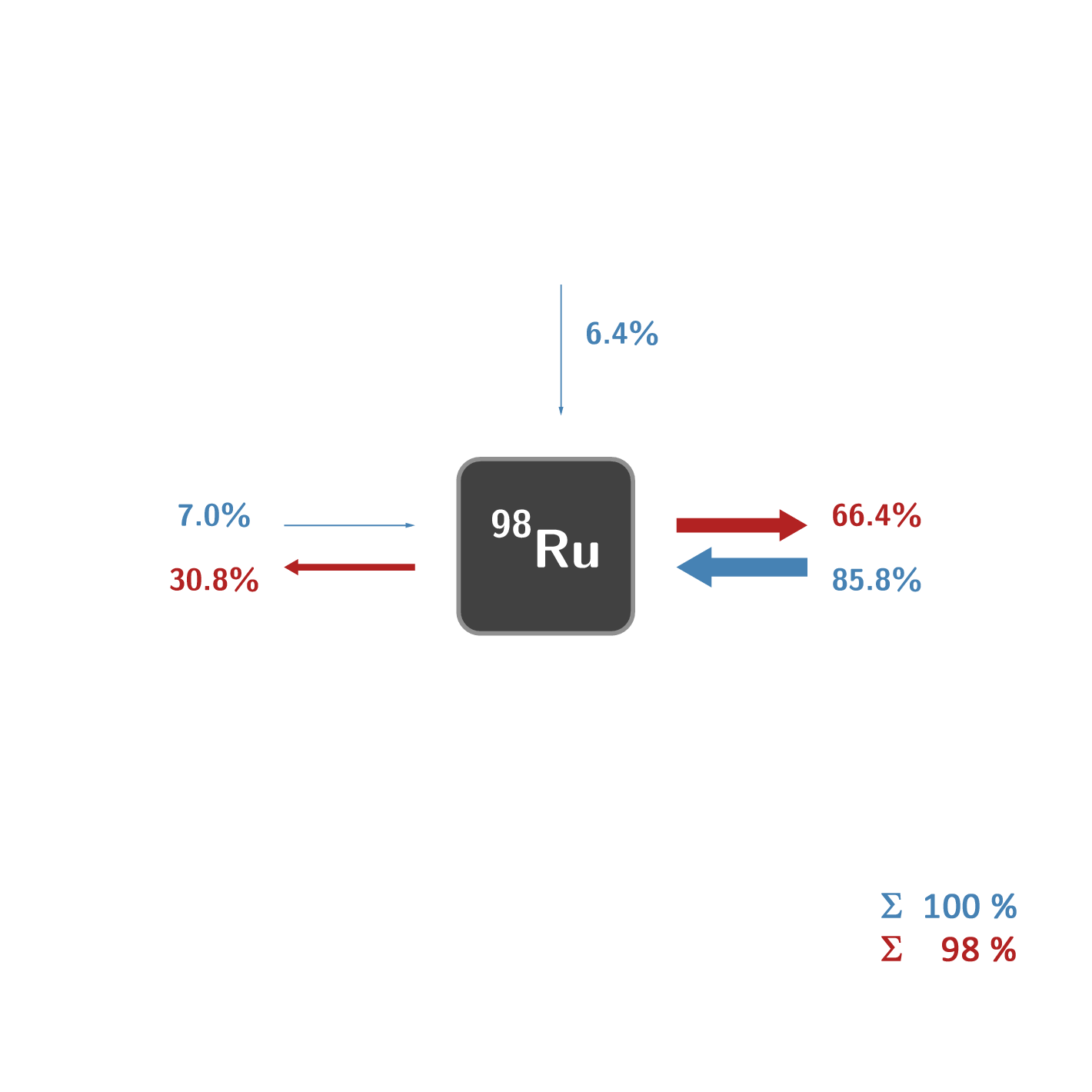}}}
\mbox{
\subfigure
{\includegraphics[trim=150 140 80 250, clip=true,width=0.5\textwidth]{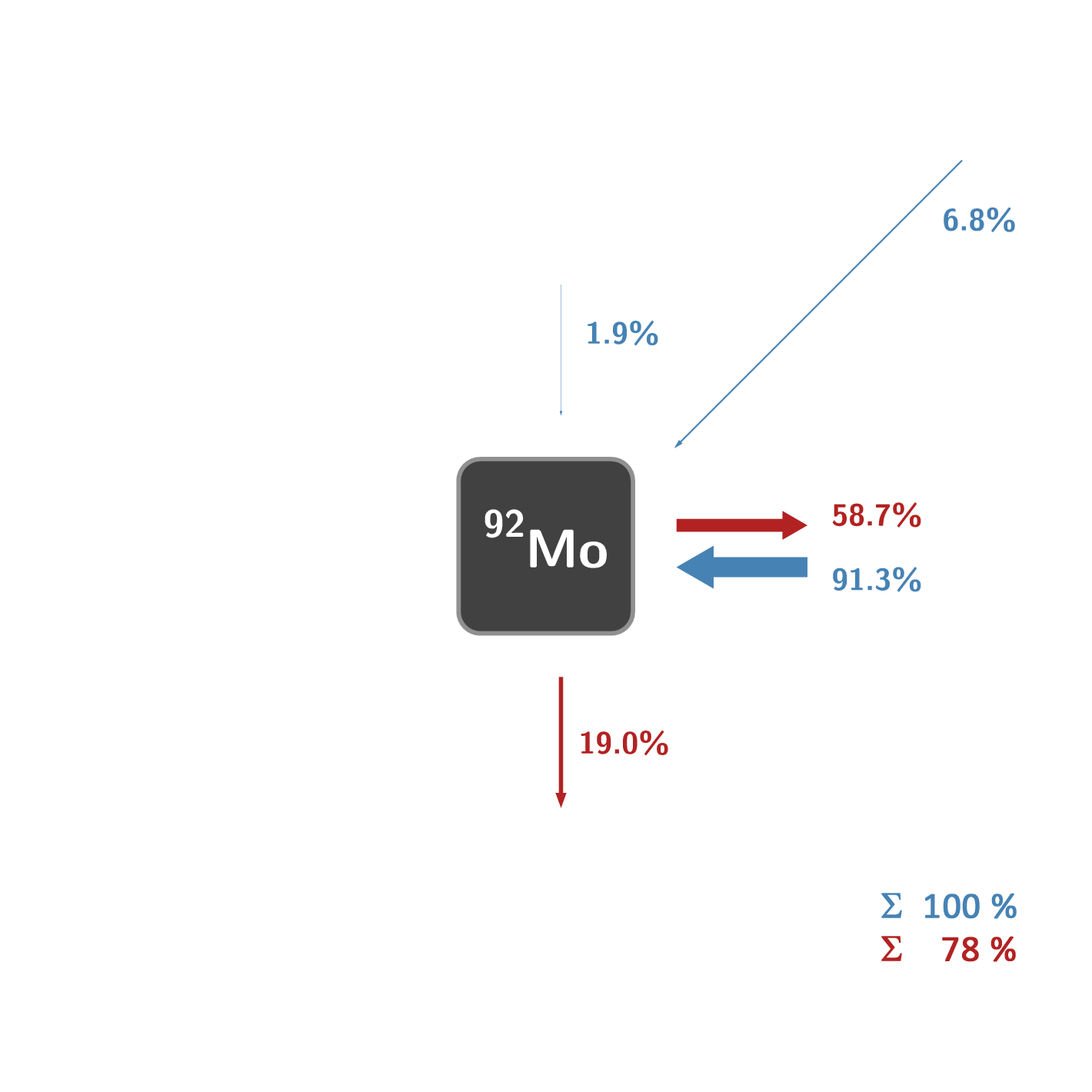}}
\subfigure
{\includegraphics[trim=150 140 80 250, clip=true,width=0.5\textwidth]{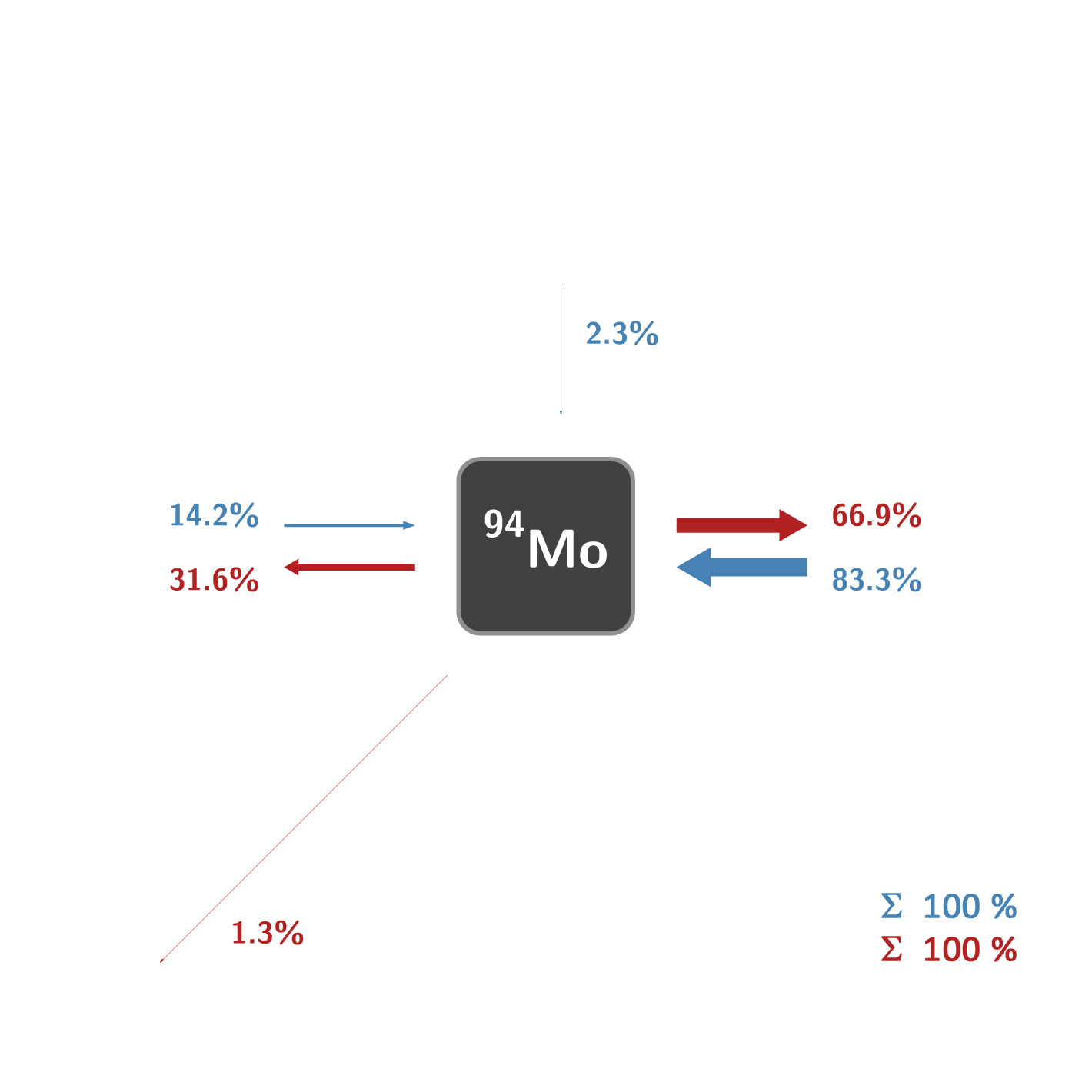}}}
\caption{Relative time-integrated fluxes producing (blue) and destroying (red) the $p$~nuclei $^{92}$Mo, $^{94}$Mo, $^{96}$Ru and $^{98}$Ru in the post-processing nucleosynthesis simulation of Hashimoto trajectory 3. The sum of all production fluxes is normalized to 100\% for each isotope, and the destruction fluxes are scaled with the same factor. Fluxes smaller than 1\% are not shown.}
\label{fig:Hash03RefFluxes} 
\end{figure}

\begin{figure}
\centering \mbox{
\subfigure
{\includegraphics[trim=150 140 80 400, clip=true,width=0.5\textwidth]{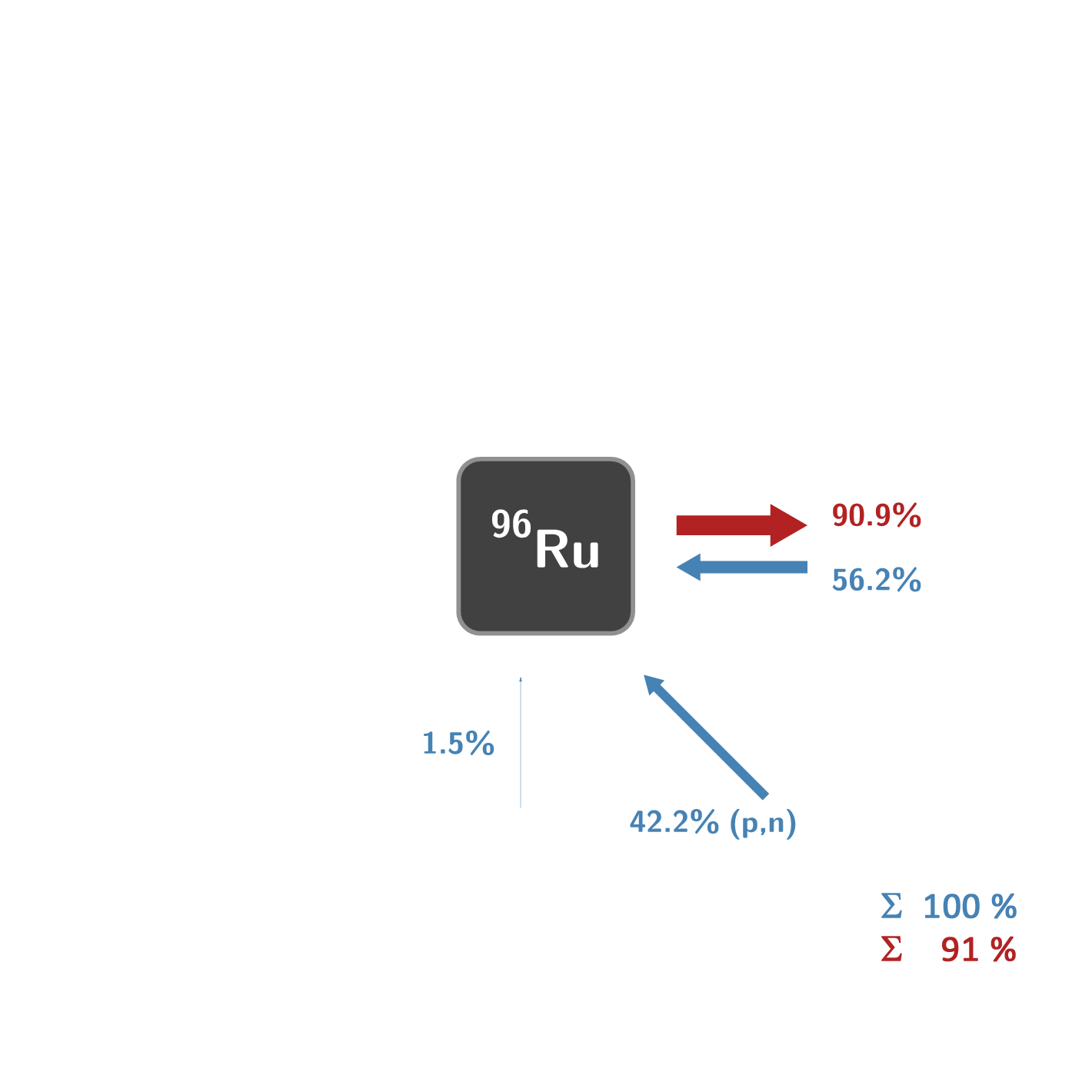}}
\subfigure
{\includegraphics[trim=150 140 80 400, clip=true,width=0.5\textwidth]{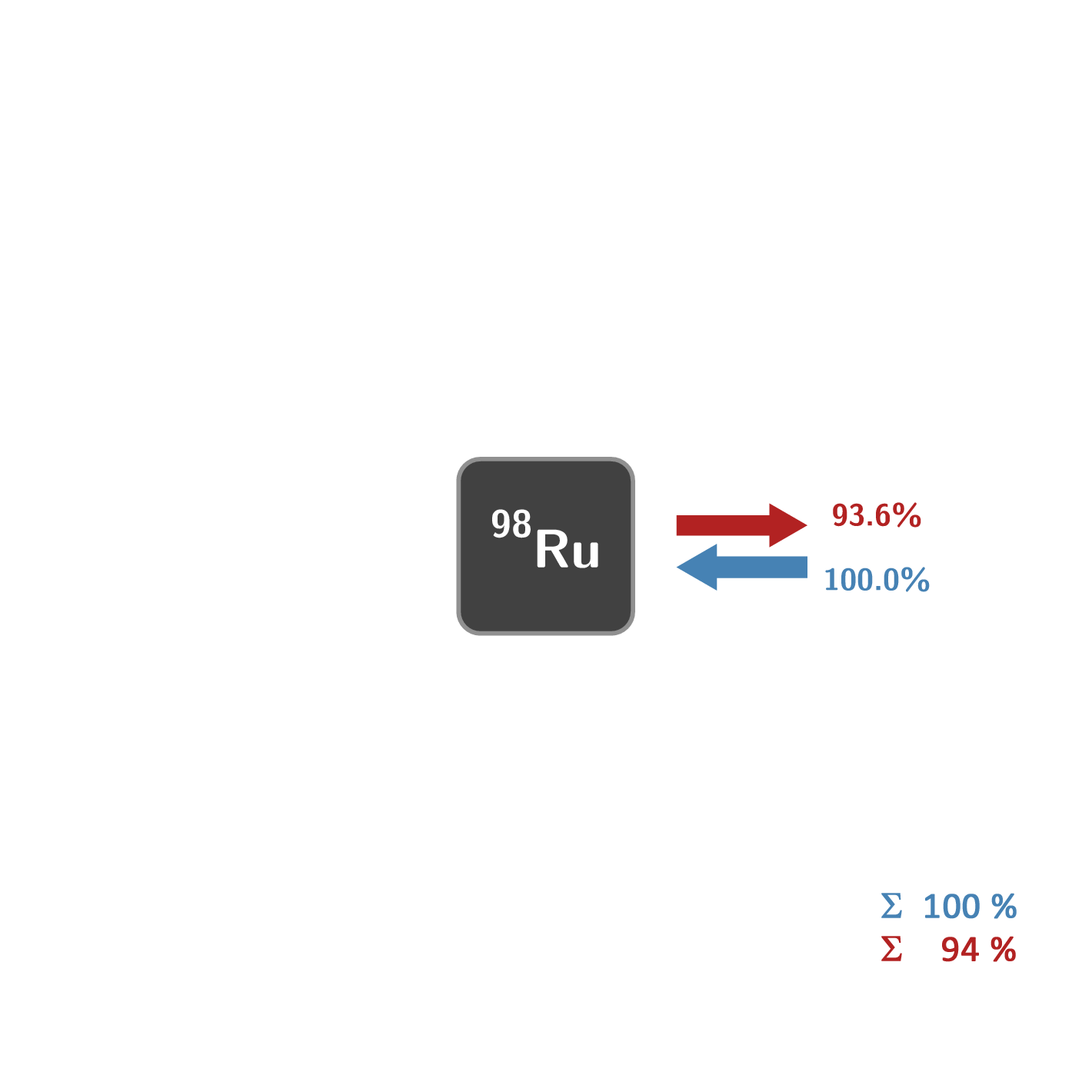}}}
\mbox{
\subfigure
{\includegraphics[trim=150 140 80 500, clip=true,width=0.5\textwidth]{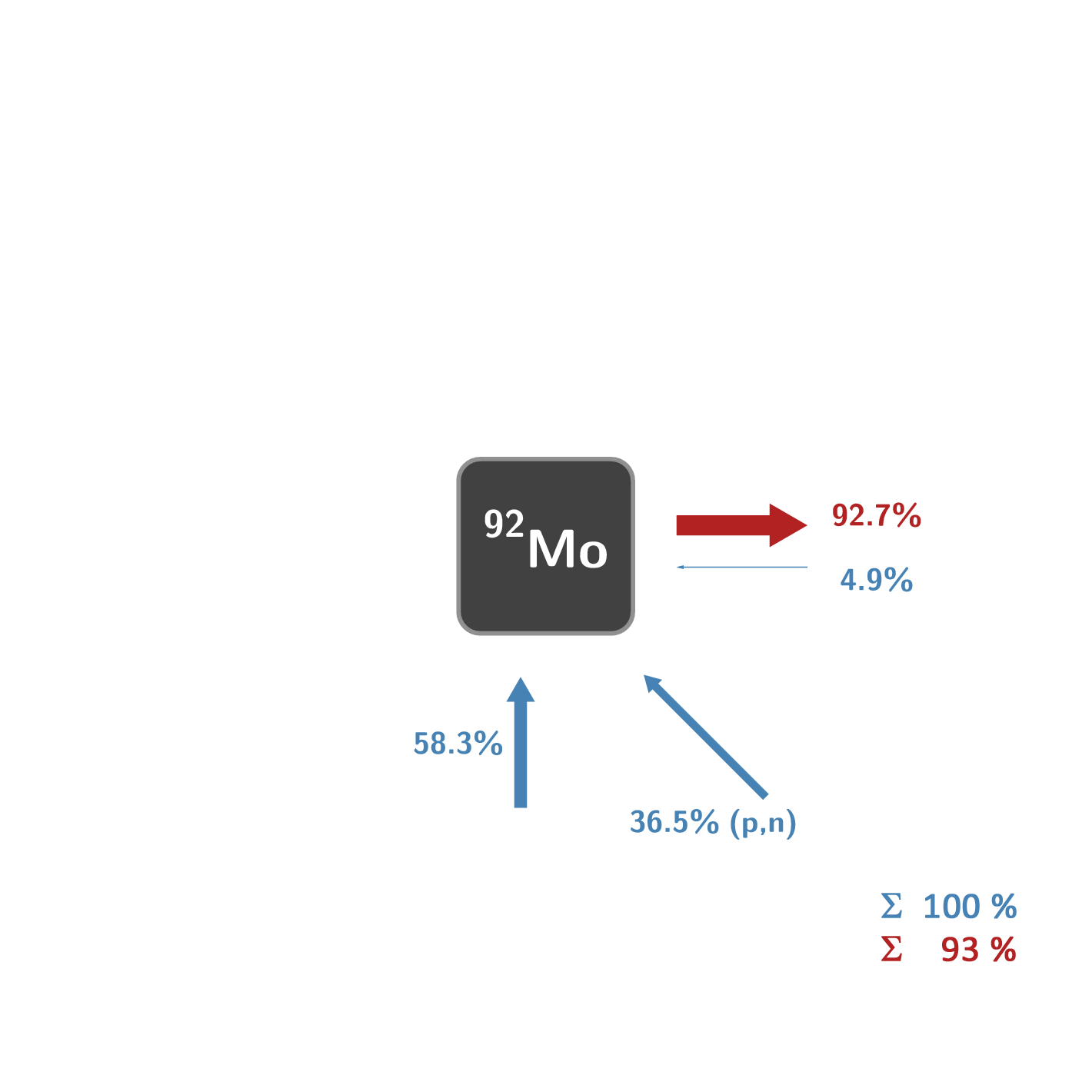}}
\subfigure
{\includegraphics[trim=150 140 80 500, clip=true,width=0.5\textwidth]{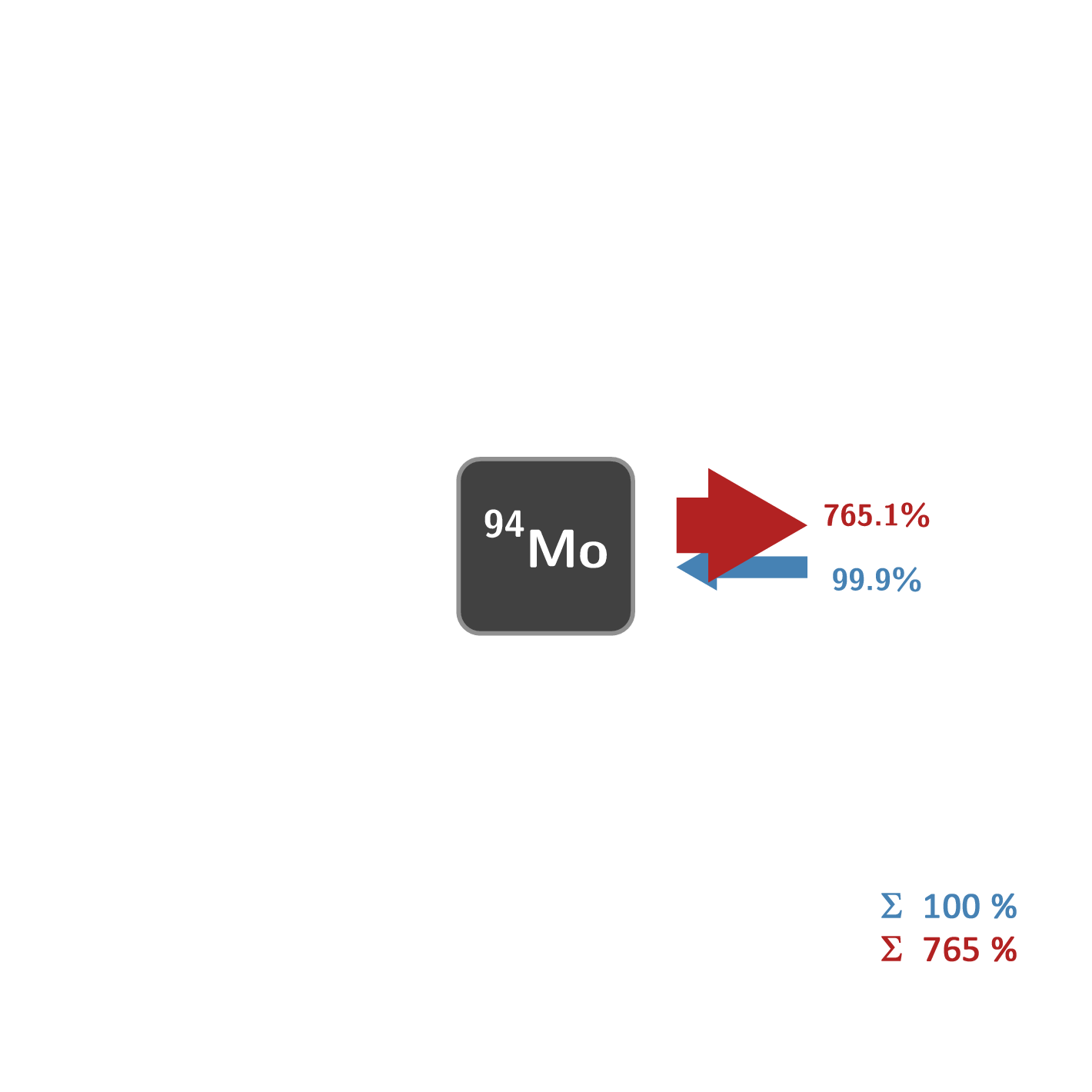}}}
\caption{The same as Fig.~\ref{fig:Hash03RefFluxes}, for trajectory 6.}
\label{fig:Hash06RefFluxes} 
\end{figure}

Fig.~\ref{fig:Hash03RefFluxes} shows the time-integrated relative nucleosynthesis fluxes for the $p$ nuclei \isotope[92]{Mo}, \isotope[94]{Mo}, \isotope[96]{Ru} and \isotope[98]{Ru} for Hashimoto trajectory 3~\cite{gobel:15,expastroOnlineFluxes}. The isotopes are produced and destroyed mainly by neutron-dissociation and neutron-capture reactions in the isotopic chains. The isotopes \isotope[92]{Mo} and \isotope[96]{Ru} show a significant net production in this environment. The picture is different in the cooler environment of Hashimoto trajectory 6 (Fig.~\ref{fig:Hash06RefFluxes}). Neutron-capture reactions dominate the destruction fluxes of the Mo and Ru $p$~nuclei. Here, also neutron-dissociation reactions produce these isotopes, but proton-capture and (p,n) reactions are as or even more important for the production of \isotope[92]{Mo} and \isotope[96]{Ru}. However, these contributions only play a minor role for the production of the Mo and Ru $p$~nuclei in the CCSN model since the abundances of the isotopes obtained in this cooler environment are much lower than for the hotter environments (compare Fig.~\ref{fig:HashAbu} and the abundances in Fig.~\ref{fig:abund_and_flux_zr-cd}). 

\begin{center}
\begin{figure}
\centering
\resizebox{10cm}{!}{\rotatebox{0}{\includegraphics{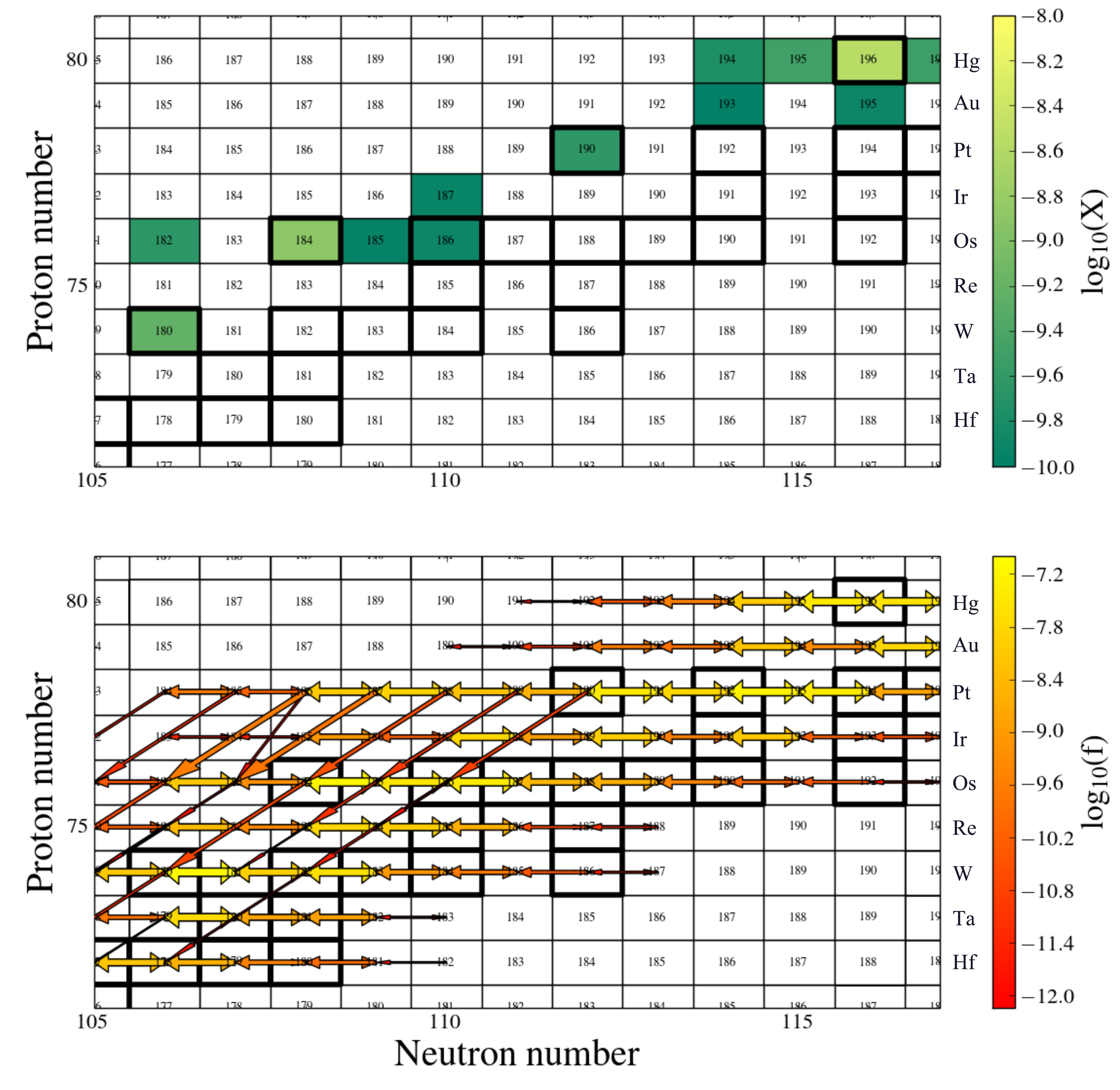}}}
\caption{Mass fraction distributions (green) and time-integrated nucleosynthesis fluxes (arrows with red to yellow color) in the mass region between Ta and Hg for Hashimoto trajectory 6. (Compare Fig.~\ref{fig:abund_and_flux_zr-cd}, and note the different axis ranges for the abundances.)}
\label{fig:abund_and_flux_ta-hg}
\end{figure}
\end{center}

\begin{figure}
\centering \mbox{
\subfigure
{\includegraphics[trim=150 300 80 500, clip=true,width=0.5\textwidth]{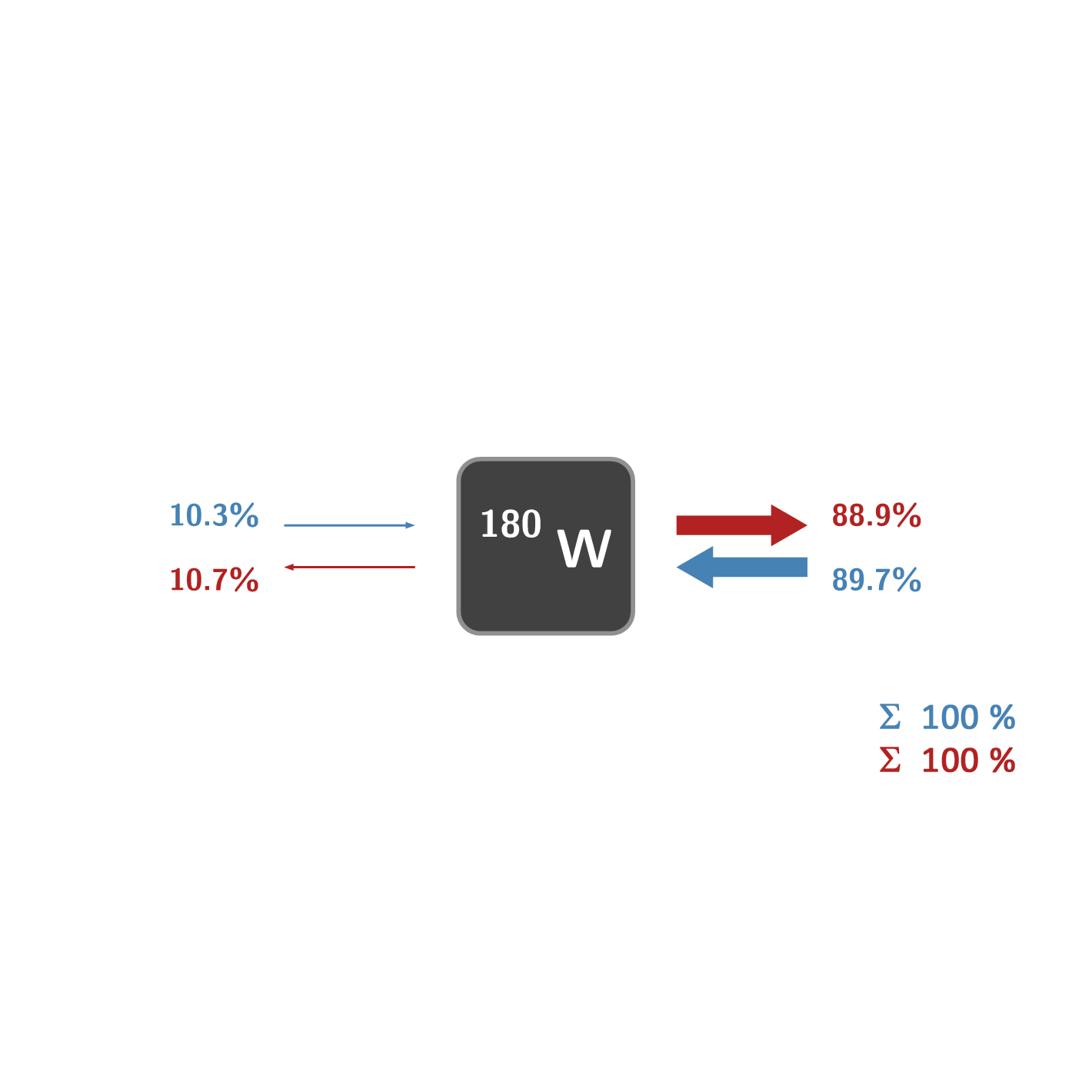}}
\subfigure
{\includegraphics[trim=150 300 80 500, clip=true,width=0.5\textwidth]{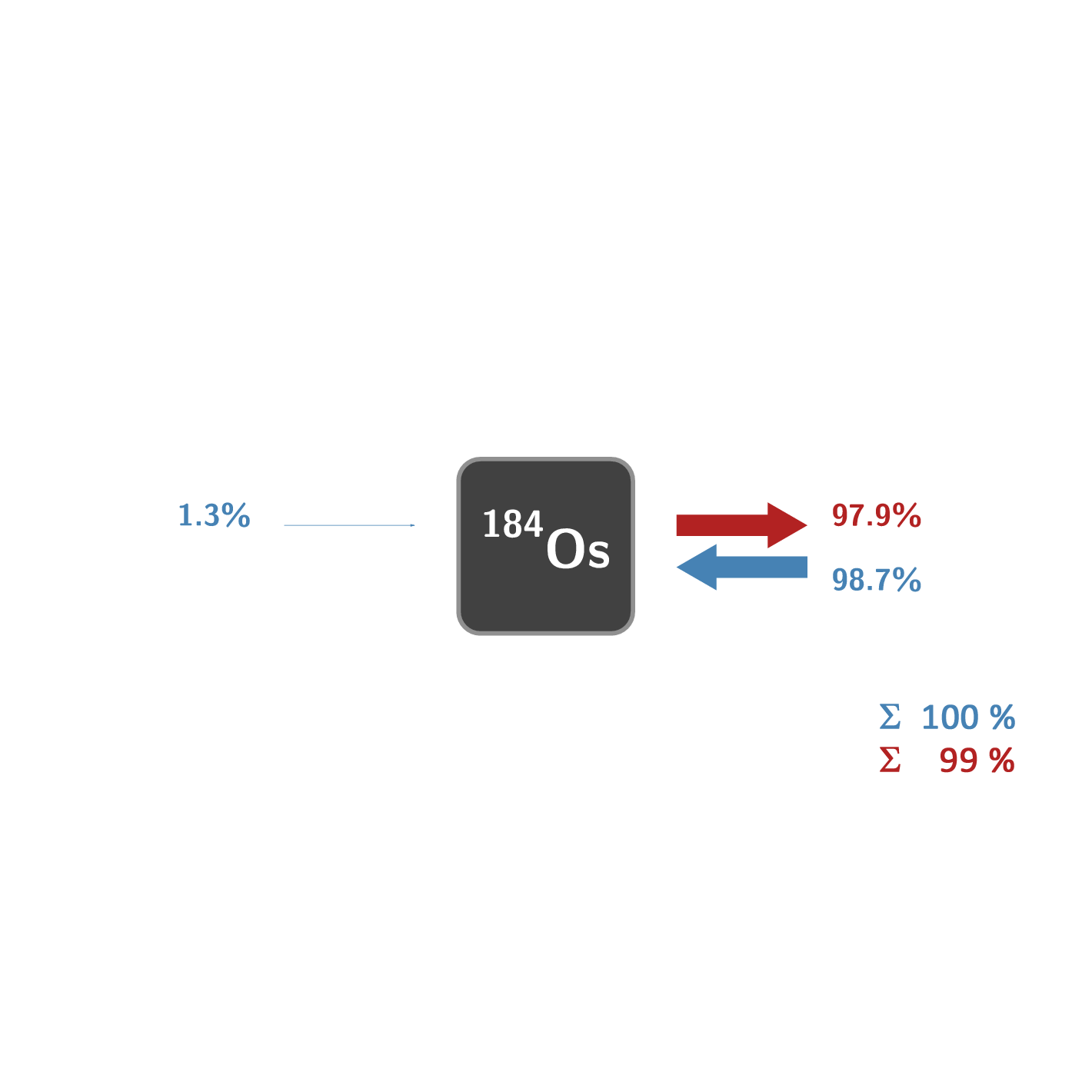}}}
\mbox{
\subfigure
{\includegraphics[trim=150 300 80 600, clip=true,width=0.5\textwidth]{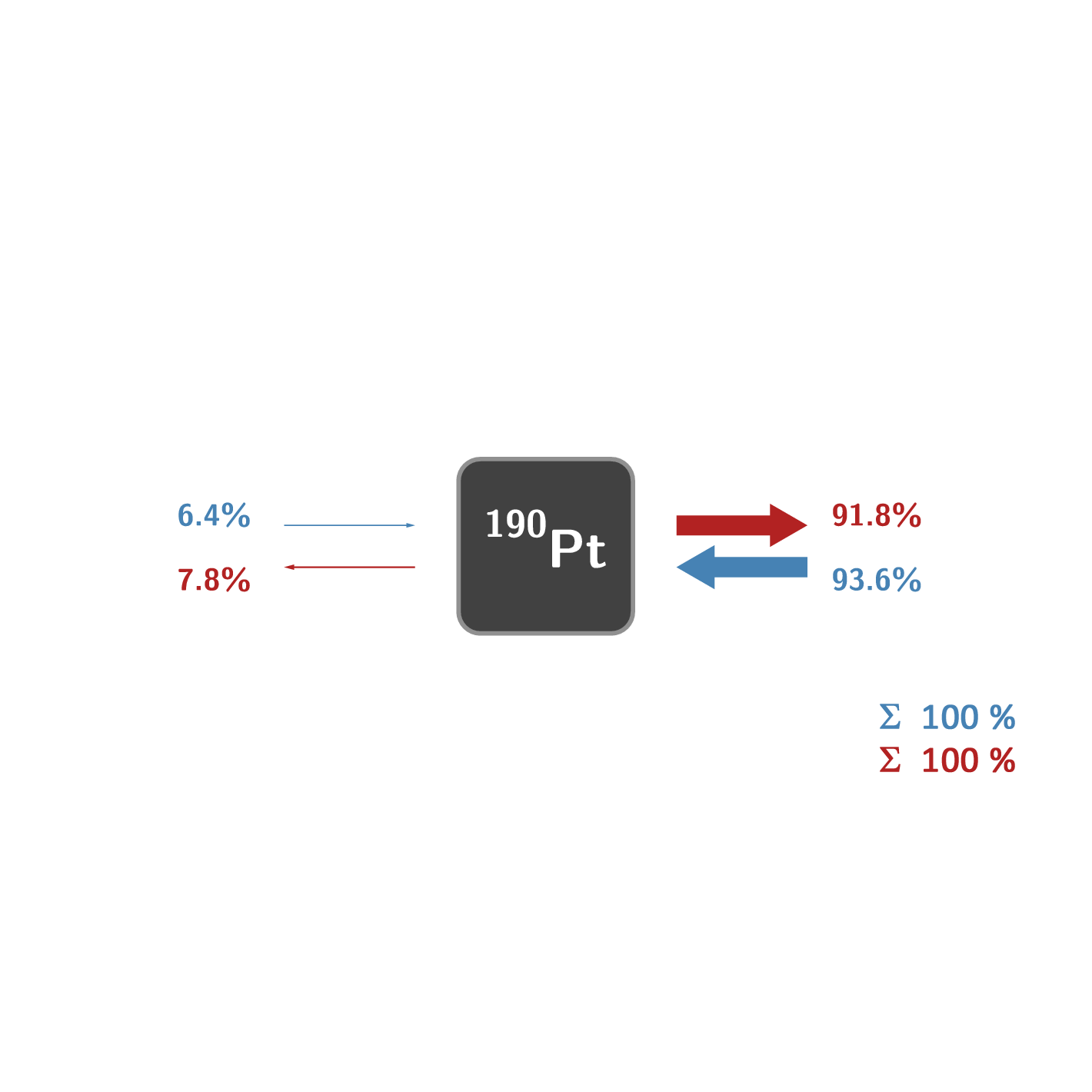}}
\subfigure
{\includegraphics[trim=150 300 80 600, clip=true,width=0.5\textwidth]{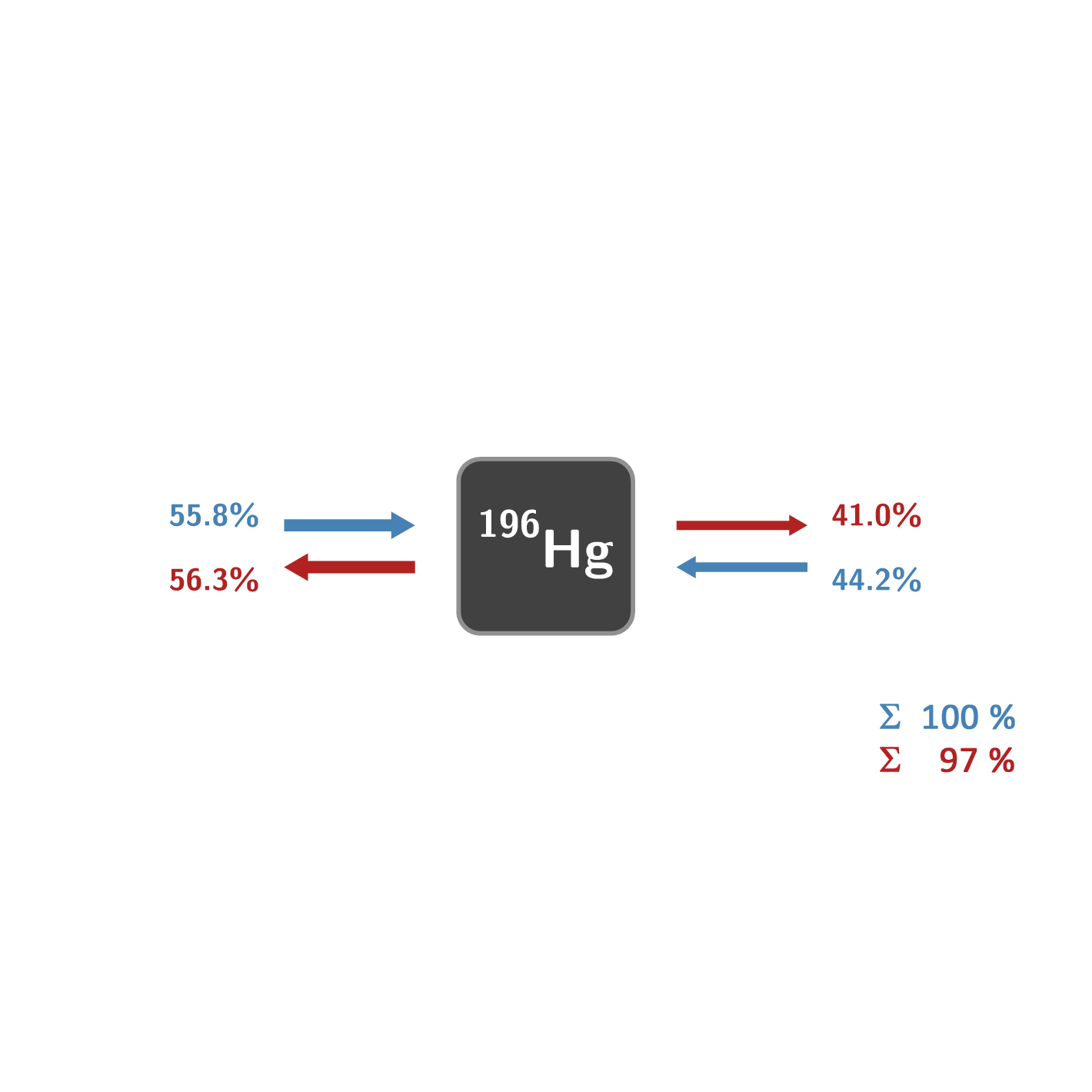}}}
\caption{The same as Fig.~\ref{fig:Hash06RefFluxes}, for the $p$ nuclei $^{180}$W, $^{184}$Os, $^{190}$Pt and $^{196}$Hg.}
\label{fig:Hash06RefFluxesHeavy} 
\end{figure}

The absolute nucleosynthesis fluxes and the resulting abundances in the mass region between Ta and Hg, where the $p$ nuclei $^{180}$Ta, $^{180}$W, $^{184}$Os, $^{190}$Pt and $^{196}$Hg are located, are shown in Fig.~\ref{fig:abund_and_flux_ta-hg}. Neutron-dissociation and neutron-capture reactions dominate the fluxes (see also Fig.~\ref{fig:Hash06RefFluxesHeavy}). $\alpha$-dissociations on proton-rich nuclei feed lighter elements. In total, few proton-rich isotopes are produced in this environment. The final abundances are much lower compared to lighter nuclei (compare Figs.~\ref{fig:abund_and_flux_zr-cd}~and~\ref{fig:abund_and_flux_ta-hg}).  

\subsection{Underproduction of the $p$ nuclei}

Refs.~\cite{rayet:90,prantzos:90,rayet:95} obtained two crucial results for the $\gamma$ process in CCSNe: i) the isotope \isotope[16]{O} and the $p$ nuclei are underproduced by about a factor of four compared to the Solar System abundances; ii) while the $\gamma$-process abundance distribution is qualitatively compatible with the solar distribution, the isotopes \isotope[92,94]{Mo} and \isotope[96,98]{Ru} are underproduced by one order of magnitude or more compared to the other $p$ nuclei. 

The first conclusion was not based on detailed galactical chemical evolution (GCE) simulations since GCE simulations to properly compare the $\gamma$-process production in CCSNe and the solar abundances are unavailable at the moment~\cite{rauscher:13}. 
A direct comparison between the production of $p$ nuclei and \isotope[16]{O} was used instead to highlight this problem. Also, the second conclusion is still a major matter of debate today, and one of the main drivers in the nuclear astrophysics community to better understand the nucleosynthesis in the Mo-Ru region. According to Refs.~\cite{rauscher:06,rapp:06}, these issues cannot be resolved by nuclear physics uncertainties.

\subsection{Impact of seed distributions and CCSN uncertainties}

\subsubsection{Fusion reactions and the weak $s$ process}

The evolution of the massive star up to its fate as a CCSN determines the seed distribution for the $p$~process. Convective C-burning shells are evolving in the O/Ne-rich layers of the star. Oxygen is the most abundant element in these layers, which has been produced at the end of the previous convective He-burning core~\cite{woosley:95}. The second most abundant isotope is \isotope[20]{Ne}, a direct product of carbon fusion together with \isotope[23]{Na}~\cite{thielemann:85}. This region is also enriched with heavy $s$-process elements. Neutrons are produced by the reaction \isotope[22]{Ne}($\alpha$,n)\isotope[25]{Mg} in the convective He-burning core and later in the C shell. Neutron capture reactions on Fe seeds produce the weak $s$-process elements in the mass region 60$<$A$<$90~\cite{raiteri:91a,raiteri:91b,the:07,pignatari:10,kaeppeler:11}. In baseline massive star models no significant production is obtained beyond A$\sim$90. 

However, seed distributions up the heavy elements Pb and Bi are needed to produce the $p$ nuclei by photodisintegration reactions during the $\gamma$~process. For a massive star, the abundances of the elements with $A>$90 are basically given by the pristine stellar composition. The freshly produced material by fusion reactions and the weak $s$~process 
are not the main seeds for the $p$ nuclei in massive stars. 
Hence, the uncertainties related to these s-process scenarios do not strongly affect the outcome of the $\gamma$-process simulations~\cite{raiteri:91b,pignatari:10}.

Effects of single rates on the abundances obtained from a massive star have been investigated. 
Ref.~\cite{costa:00} argued that a larger \isotope[22]{Ne}($\alpha$,n)\isotope[25]{Mg} rate (within the upper limit of Ref.~\cite{angulo:99}) could increase the $s$-process yields in massive stars at the Sr peak and beyond. However, this result has not been confirmed~\cite{heger:02}. Also, such a high rate for $\alpha$-capture on \isotope[22]{Ne} has not been confirmed by more recent experiments~\cite{longland:12,talwar:15}. Nucleosynthesis results would disagree with present weak and main $s$-process component predictions in the Solar System abundance distribution.

The $p$ nuclei, in particular the light $p$~nuclei \isotope[92,94]{Mo} and \isotope[96,98]{Ru}, are underproduced in stellar simulations using the described seed abundances. Ref.~\cite{arnould:92} showed that the artificial increase of the $s$-process abundances for A $\geq$ 90 solves the Mo-Ru puzzle. This was confirmed by recent stellar calculations by Ref.~\cite{pignatari:13}. An enhanced \isotope[12]{C}+\isotope[12]{C} fusion reaction rate compared to Ref.~\cite{caughlan:88} affects the stellar evolution structure and enhances the final $s$-process production, due to the additional contribution of the \isotope[13]{C}($\alpha$,n)\isotope[16]{O} reaction rate. 
As a consequence, the Mo and Ru $p$~nuclei production reaches the level of other $p$~nuclei, and the overall $\gamma$-process strongly increases 

Therefore, uncertainties affecting the $s$-process seeds production might also be relevant for the $\gamma$-process.

\subsubsection{Metallicity and stellar rotation}

The $\gamma$-process yields depend on the initial metallicity of the star. The production of the weak $s$ process nuclei is secondary and decreases if the initial concentration of metals in the star decreases~\cite{raiteri:92,baraffe:92,pignatari:08}. Beyond A$\sim$90, the seeds abundances are similar to the initial abundances, and therefore secondary by definition. As a consequence, also the $\gamma$ process in CCSNe is a secondary process. 

According to Ref.~\cite{tinsley:80}, a secondary isotope should be overproduced by a factor of two  at solar metallicity compared to a primary isotope if they are fully synthesized by the same astrophysical source. Hence, the secondary $p$ nuclei should be overproduced by a factor of two compared to primary isotopes like \isotope[16]{O}. This challenges the $\gamma$ process in present CCSNe models to explain all the $p$ nuclei in the Solar System.

Recently, Ref.~\cite{chieffi:15} have obtained models of fast rotating massive stars at metallicity lower than solar with $s$-process production efficient up to Pb, due to the activation of the \isotope[13]{C}($\alpha$,n)\isotope[16]{O} reaction rate. This result is not confirmed by other models~\cite{frischknecht:16}, but this scenario may have important implications for the $\gamma$-process production at low metallicities, since these enhanced seeds would also greatly enhance the $\gamma$-process products. More detailed studies are required for the $\gamma$ process in fast-rotating massive stars.

\subsubsection{Progenitor mass and CCSN explosions}

The first comprehensive studies about the $\gamma$~process in CCSNe assumed that the final results are independent of the initial mass of the CCSN progenitors~\cite{rayet:95}. This may be explained by the well-constrained thermodynamics conditions triggering the $\gamma$ process that are naturally obtained in CCSNe. Also, the weak $s$-process seeds are not expected to drastically change their main properties by considering a 15$\msun$ star or a 60$\msun$ star, even if results may be significantly different for a detailed $s$-process analysis~\cite{prantzos:90,kaeppeler:94}. In particular, the present $\gamma$-process seeds are robust if the dominant neutron source for the weak $s$ process is the reaction \isotope[22]{Ne}($\alpha$,n)\isotope[25]{Mg}. 

\begin{figure}
\centering
\resizebox{11cm}{!}{\rotatebox{0}{\includegraphics{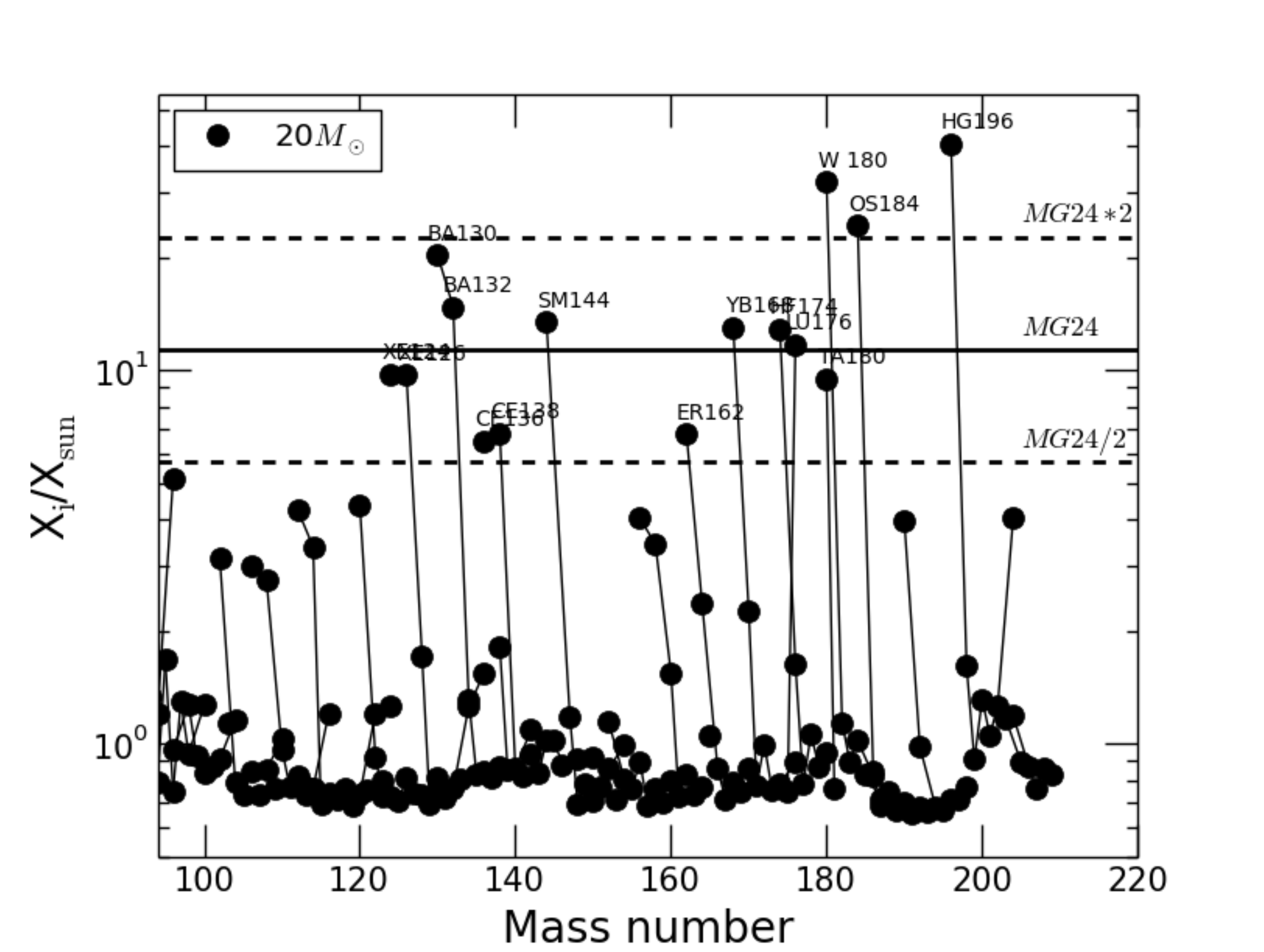}}}
\resizebox{11cm}{!}{\rotatebox{0}{\includegraphics{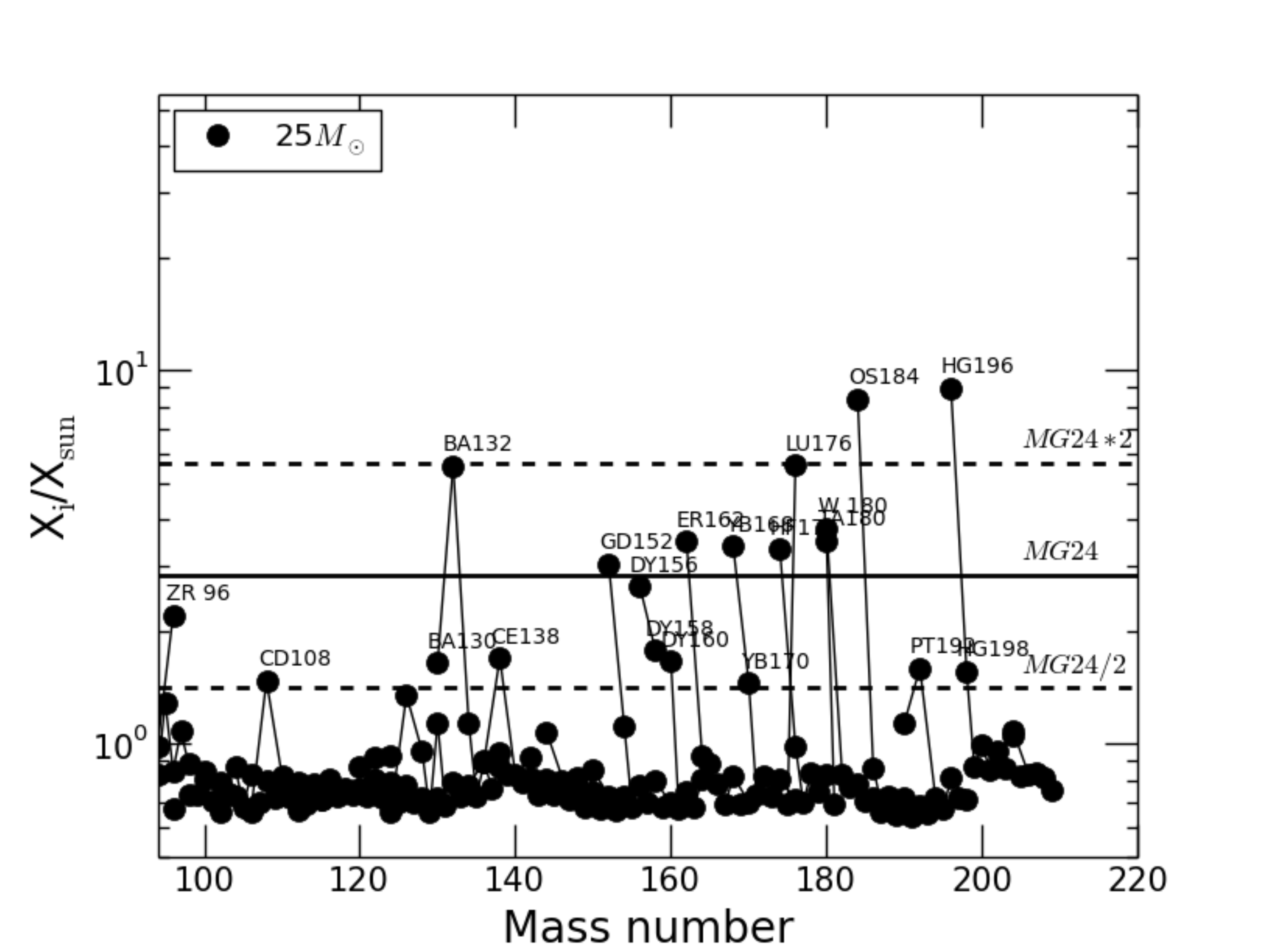}}}
\caption{Final isotopic overproduction factors for the 20 (upper panel) and 25$\msun$ (lower panel) stars in the mass region A $>$ 95. The stellar calculations are from Ref.~\cite{pignatari:13a}, and used solar metallicity as the seed distribution. The production factor of \isotope[24]{Mg}, divided and multiplied by a factor of two are also reported (continuous and dashed lines). We label the isotopes with production factors larger than \isotope[24]{Mg} divided by 2. Among those, different $p$~nuclei can be identified.}
\label{fig:pprocess_mass}
\end{figure}

Ref.~\cite{rauscher:02} showed that the $\gamma$ process products can show relevant differences from one case to the other considering a grid of stellar models. In Ref.~\cite{pignatari:13a} we discussed instead the large impact of the CCSN explosion. Fig.~\ref{fig:pprocess_mass} shows the isotopic production factors for CCSN models with initial masses 20\msun\ and 25\msun\ and solar metallicity for A$>$95, just beyond the mass region where the weak $s$~process is contributing. Most of the highlighted species are $p$ nuclei.

\isotope[16]{O} has been used as reference isotope to quantify the efficiency of the $\gamma$ process without consistent GCE calculations to study the production of $p$ nuclei. However, the oxygen abundance in the Sun is still uncertain. Last results based on modern solar atmosphere models (e.g., Refs.~\cite{asplund:09,caffau:15}) calculate a much smaller oxygen abundance compared to older abundances~\cite{anders:89,grevesse:93}, and a final answer is not available yet. Ref~\cite{rauscher:13} highlighted that the oxygen abundance is crucial since its uncertainty is directly propagated to the analysis of the $p$-nuclei production. Therefore, we use \isotope[24]{Mg} as a reference isotope instead of \isotope[16]{O}. In general, \isotope[24]{Mg} has similar properties as \isotope[16]{O}: it is mostly produced in massive stars, it is primary, it is not much affected by the CCSN explosion energy, and it is made during the pre-explosive hydrostatic phase~\cite{thielemann:96}. The only difference is that \isotope[24]{Mg} is mainly a C-burning product, while \isotope[16]{O} is made mostly during the He-burning phase.

In Fig.~\ref{fig:pprocess_mass}, the general underproduction of $p$ nuclei compared to the reference isotope obtained by early works is not so evident. This was already shown in some models by Ref.~\cite{rauscher:02}. Physics reasons may play some role, but a better zone resolution of more modern stellar models has also improved the quality of the $\gamma$-process yields~\cite{rauscher:15}. The figure also shows a variation in the order of a factor of four between different isotopes. Some of them show a potentially full production according to the criteria mentioned above (a factor of two larger production than the primary \isotope[24]{Mg}), while others are still underproduced. 

Finally, the 25\msun\ model shows a much lower production for the $p$ nuclei and \isotope[24]{Mg} compared to the 20\msun\ model. The variation results from the CCSN recipe used in these models, where the fallback mass after the explosion tends to increase with the progenitor mass~\cite{fryer:12,pignatari:13a}. Only part of the $\gamma$~process and the \isotope[24]{Mg} rich material is ejected in the 25\msun\ model, while they are ejected completely in the 20\msun\ model.


The uncertainties in the SN explosion can significantly affect the $\gamma$-process yields. The fallback parametrization in one-dimensional CCSN models may change significantly depending on the prescription used~\cite{fryer:12,ugliano:12}. Furthermore, the multi-dimensional nature of the CCSN explosion (e.g., Refs.~\cite{janka:12,burrows:13,hix:14,wongwathanarat:15}) makes less constraining the use of \isotope[24]{Mg} or \isotope[16]{O} as a reference. They are only marginally affected by the SN explosion, while the $\gamma$ process (or at least a good part of it) is an explosive product.

Therefore, different CCSN prescriptions and different stellar structure evolutions can drastically change the final $\gamma$-process yields. To fully appreciate the astrophysical impact of this uncertainty and study the relevance for the Solar System abundances it is mandatory to perform GCE simulations of the $p$ nuclei.

\section{The production of $p$ nuclei in thermonuclear supernovae}
\label{pprocess-snia}

Thermonuclear explosions of Chandrasekhar-mass ($M_{\mathrm{Ch}}$) carbon-oxygen white dwarfs (hereafter CO-WD) are favored as the progenitor for a majority of Type Ia supernova (SNIa)~\cite{hoflich:96,fink:07}. The mass of the WD could reach the $M_{\mathrm{Ch}}$ by several evolutionary paths: by a mass transfer from a giant/main sequence binary companion (single-degenerate scenario~\cite{nomoto:82}), or by merging with a degenerate WD binary companion (double-degenerate scenario~\cite{iben:84,ropke:12}). In addition, sub-$M_{\mathrm{Ch}}$ explosions have also been investigated~\cite{arnett:94}. 

Typically, the WD material is converted to iron-peak elements in the explosion process (Fe, Ni and neighboring elements which form a prominent peak in the Solar System abundance profile), and a smaller fraction of intermediate-mass elements. SNe~Ia have a characteristic light curve, which is triggered by the decay of the produced radioactive nuclei. The spectra show no signs of hydrogen and helium. Intermediate mass elements (such as Si, S, and Ca), which constitute the outer layers of the star, are evident near maximum luminosity. At late times, the spectra are dominated by heavy elements (like Ni, Co, and Fe), which are produced during the explosion near the core~\cite{hillebrandt:00}. However, the observed heterogeneity of SNe~Ia suggests multiple progenitors and explosion mechanisms. It is unclear how the proposed models contribute to the observed SNe~Ia. (See Refs.~\cite{hillebrandt:00,hillebrandt:13} for detailed reviews.)

The models of a thermonuclear disruption of a white dwarf are able to fit the observed light curves and spectra well. However, the evolution of massive WDs to explosion is very uncertain, leaving room for some diversity in the allowed set of initial conditions (such as the temperature profile at ignition). Furthermore, the physics of thermonuclear burning in degenerate matter is complex and not fully understood.

\subsection{Nucleosynthesis simulations for multidimensional SNe Ia}

Our understanding of the nucleosynthesis in SNe~Ia is based on decades of spherically symmetric modeling. In recent years, detailed estimates of the nucleosynthesis yields for multidimensional explosion models have become possible. Two dimensional asymmetric simulations~\cite{livne:93,reinecke:99,ropke:09} and three dimensional simulations~\cite{reinecke:02,gamezo:03,seitenzahl:13} of exploding $M_{\mathrm{Ch}}$WD have provided new perspectives on SNIa nucleosynthesis calculations.  

\begin{figure}
\centerline{\includegraphics[width=13cm]{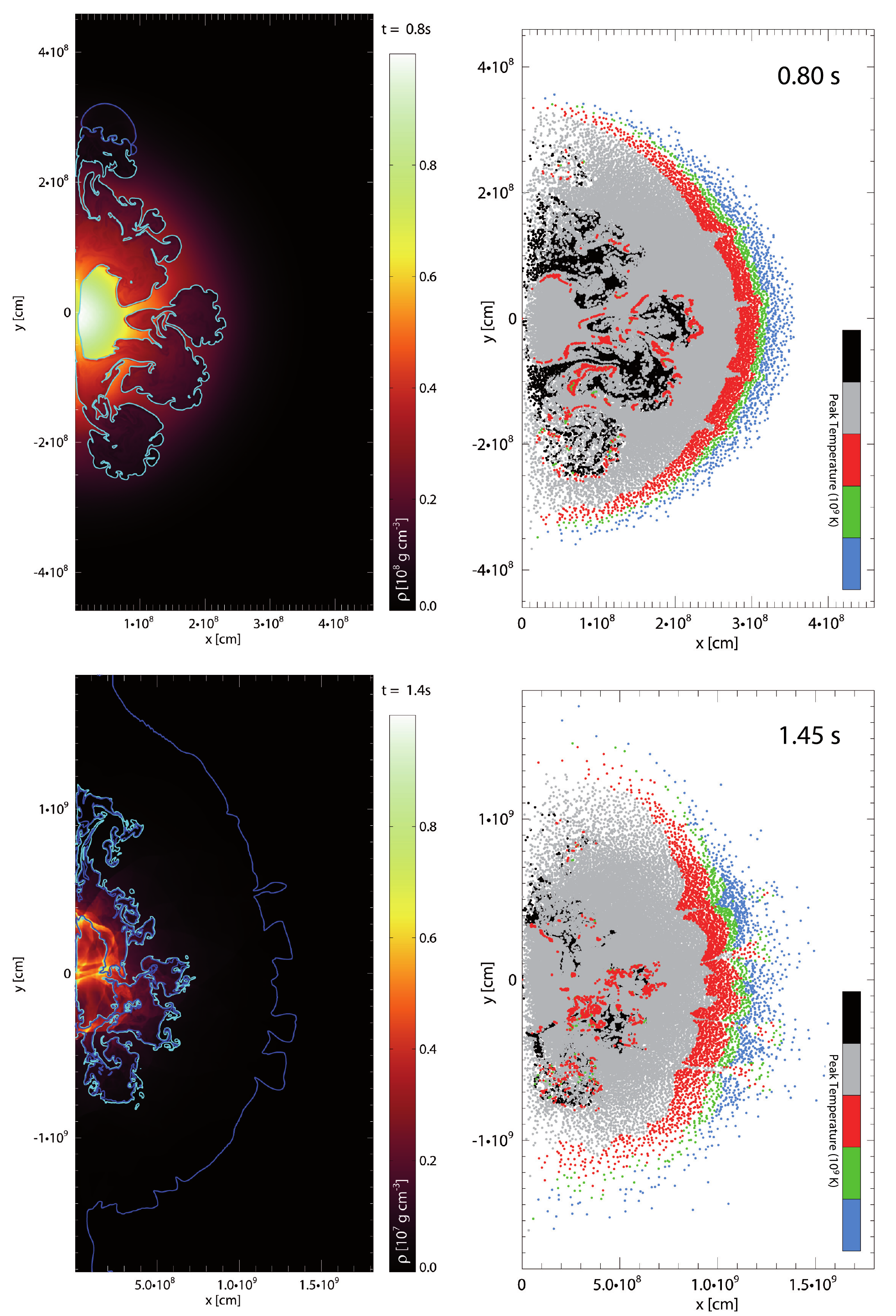}}
\caption{Snapshots of the 2D model (deflagration-to-detonation transition model, CO-WD structure~\cite{dominguez:01} with $Z$~=~0.02 and a progenitor mass of $M$~=~1.5$\msun$.) 0.8~s and 1.45~s after ignition. The hydrodynamic evolution is plotted on the left, the density being color-coded. The cyan contour indicates the position of the deflagration flame while the blue contour shows the detonation front. The right panel displays the distribution of the 51,200 tracer particles used in this simulation. Their maximum temperature is color-coded.}
\label{fig:2Dmodelsnapshots}
\end{figure}

Coupled systems of hydrodynamic and nuclear kinetic equations can only be solved fully in one-dimensional stellar codes due to CPU time and memory limitations. In multi-dimensional codes, the hydrodynamic simulation includes only the nuclear reactions relevant for the energy production. The complete nucleosynthesis network calculation is relegated to a post-processing approach~\cite{travaglio:04,travaglio:11,travaglio:15}. The necessary temperature and density profiles for the post-processing step are obtained by the tracer particle method. A $Lagrangian$ $component$ is added to the Eulerian scheme in the form of tracer (or marker) particles. The tracer particles correspond to fluid mass elements and are passively advected with the flow in course of the Eulerian calculation. Shortly, tracer particles are a Lagrangian component in an Eulerian grid code. Masses and positions are assigned to the tracers in such a way that a density profile reconstructed from their distribution resembles that of the underlying star. During the hydrodynamical simulation, they record the history of thermodynamic conditions along their path (an example is shown in Fig.~\ref{fig:2Dmodelsnapshots}~\cite{travaglio:11}). 

The resolution, i.e. the number of tracer particles filling the volume of the star, has to be high enough to get accurate nucleosynthesis results. Ref.~\cite{seitenzahl:10} performed a resolution study for a 2D SNIa simulation and demonstrated that almost all isotopes up to iron-peak with abundances higher than about 10$^{-5}$ are reproduced with an accuracy better than 5\%  with 80$^2$ tracer particles. More recently, Ref.~\cite{seitenzahl:13} calculated tracer particle nucleosynthesis for different 3D SNIa models. The authors extrapolated the tracer resolution and yield convergence study from 2D to 3D, and claimed that one million tracer particles (100 per axis) are sufficient to reliably predict the yields for the most abundant nuclides. A detailed discussion regarding nucleosynthesis results for nuclei up iron-peak species can be found in Refs.~\cite{thielemann:86,timmes:03,travaglio:04,travaglio:05,seitenzahl:10,seitenzahl:13}. In this section, we will discuss in detail the production of $p$~nuclei in SNe~Ia.

\subsection{Seed distributions}

Relevant $\gamma$-process nucleosynthesis occurs in SNe~Ia only if there is a prior $s$-process enrichment. It is therefore essential to determine the source of the $s$-process enrichment in the exploding WD. In the single-degenerate progenitor model assumed here, there are two sources of $s$-process enrichment: (1) Thermal pulses occur during the AGB phase leading to the formation of the WD. The $s$-process isotopes are produced during this phase (TP-AGB phase, see e.g., Refs.~\cite{dominguez:01,herwig:05,straniero:06,karakas:14}). (2)~Thermal pulses can also occur when matter is accreted onto the WD, and enrich the matter accumulating on the WD~\cite{iben:81,iben:91,howard:93,kusakabe:11}. 

In the first scenario, the $s$-process material (0.1$\msun$) produced during the AGB phase is distributed over the WD core and mixed during the simmering phase. The resulting $p$-process abundance it is too low to account for a significant fraction of the solar $p$~abundances.

\begin{figure}
\includegraphics[width=\textwidth]{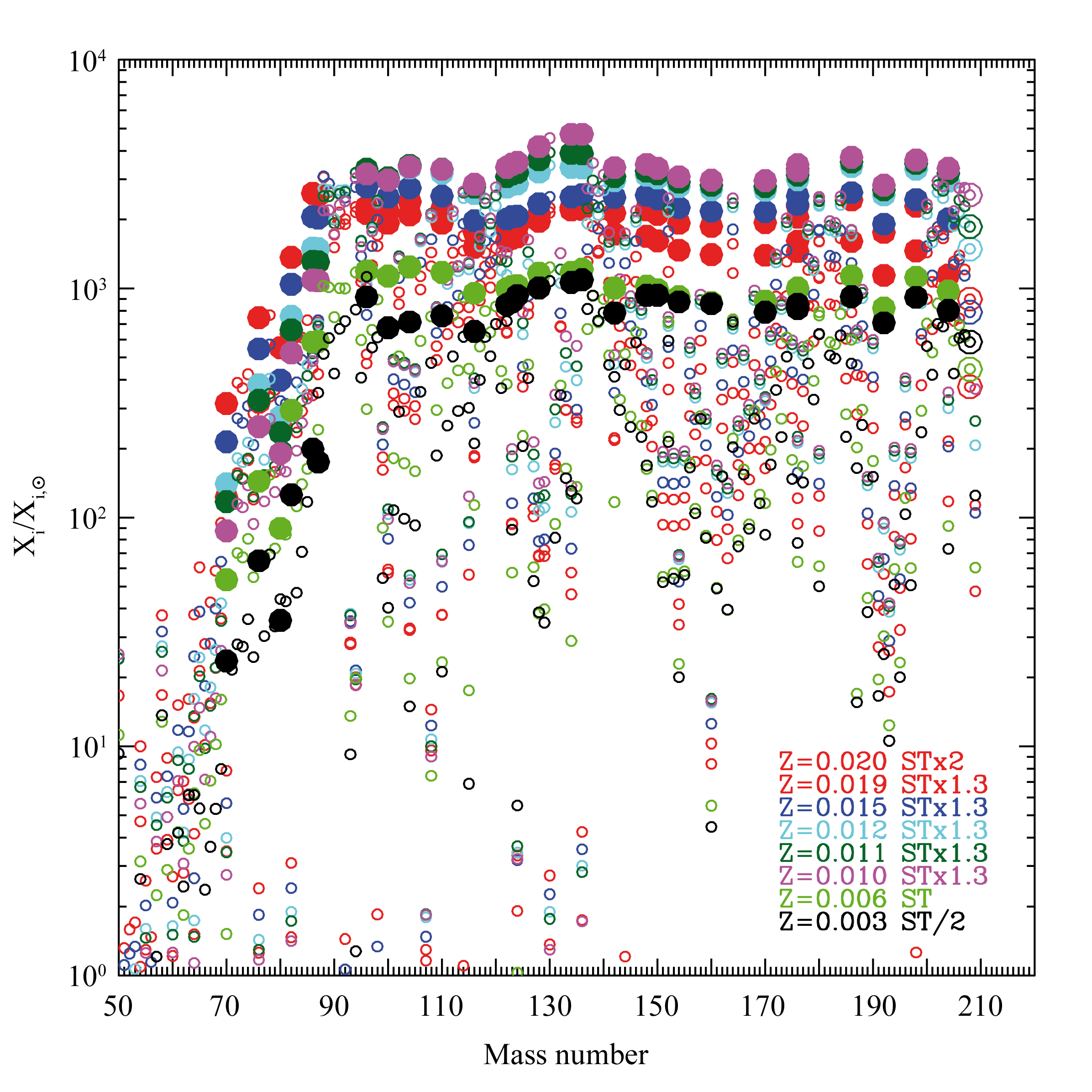}
\caption{Distribution of $s$-seed abundances relative to solar for all $^{13}$C-pocket cases and metallicities covered, used for our Galactic chemical evolution calculation. The \isotope[13]{C} abundance was varied as multiples of the standard value of about 4$\times$10$^{-6}\msun$ (Ref.~\cite{gallino:98}, ST case). The results are shown for different metallicities: $Z =$ 0.02 (red), $Z =$ 0.019 (brown), $Z =$ 0.015 (blue), $Z =$ 0.012 (cyan), $Z =$ 0.011 (dark green), $Z =$ 0.01 (magenta), $Z =$ 0.006 (light green), and $Z =$ 0.003 (black). Filled dots are for $s$-only isotopes. The big open dot is for $^{208}$Pb. The solar values are taken from Lodders~2009~\cite{lodders:09}.}
\label{fig2} 
\end{figure}

In the second scenario, $s$-process nucleosynthesis occurs in the H-rich matter accreted by the CO-WD due to recurrent He-flashes~\cite{iben:81}. Neutrons are mainly produced by the $^{13}$C($\alpha$,n)$^{16}$O reaction. $^{13}$C is produced from hydrogen and freshly made carbon, and hence, the abundance of $^{13}$C is independent of the star's metallicity (primary nature). The available neutrons are captured by $^{56}$Fe, the abundance of which depends on the metallicity (secondary nature). The lower the metallicity, the higher is the neutron exposure. Hence, this dependence compensates for the secondary nature of the $s$~elements, and results in a flat seed abundance distribution. Fig.~\ref{fig2} shows the distributions relative to solar for different $^{13}$C-pocket cases and metallicities~\cite{travaglio:15}. The nucleosynthesis calculations with the different seeds showed that a flat $s$-seed distribution directly translates into an almost flat $p$-process distribution (within a factor of about 3). The average production factor scales linearly with the adopted level of the $s$~seeds, giving the indication for a primary origin~\cite{travaglio:15}.

Ref.~\cite{travaglio:15} also demonstrated that the $^{208}$Pb $s$~seed alone plays an important role for $p$~nuclei production since the photodisintegration chains start from the heaviest nuclei. They analyzed the dependence of all the $p$~nuclei on metallicity, identifying the isotopes with a weak (like $^{92}$Mo and $^{138}$Ba) and a strong (in particular the lightest $p$~nuclei $^{74}$Se, $^{76}$Kr, and $^{84}$Sr) dependence.

\subsection{The $\gamma$ process in SNe~Ia}

The first work analyzing the possibility of efficient photodisintegration reactions and the $\gamma$-process production in Chandrasekhar-mass SNIa explosions was published by Ref.~\cite{howard:91}. They derived the initial $s$-seed distribution from helium flashes as calculated by Ref.~\cite{howard:86}. They claim that they can reproduce the abundance pattern of all $p$~nuclei, including the light $p$~nuclei, within a factor of about three. However, they obtained an overproduction of $^{74}$Se, $^{78}$Kr, and $^{84}$Sr. The abundances of these three isotopes are very sensitive to the proton density, which the authors considered rather uncertain. They also obtained a rather low production of $^{94}$Mo and $^{96}$Ru with respect to the other light $p$~nuclei. 

Later, Refs.~\cite{goriely:02,goriely:05,arnould:06} analyzed the $p$-process production in He-detonation models for sub-Chandrasekhar mass WDs. The authors considered the $s$-process solar abundances as seeds. They found the nuclides Ca to Fe to be overabundant with respect to the $p$~nuclei (with the exception of $^{78}$Kr) by a factor of about 100. They concluded that a He detonation is not an efficient scenario to produce the bulk of $p$~nuclei in the Solar System. Ref.~\cite{kusakabe:11} presented $\gamma$-process nucleosynthesis calculations in a CO-deflagration model of SNIa, i.e., the W7 model~\cite{nomoto:84}. Similar to Ref.~\cite{howard:86}, they assumed enhanced $s$-seed distributions using the classical $s$-process analysis, testing two different mean neutron exposures $\tau_o$, a flat distribution for $\tau_o$~=~0.30~mb$^{-1}$, and a decreasing $s$-process distribution corresponding to $\tau_o$~=~0.15~mb$^{-1}$.

\begin{figure}
\centering
\includegraphics[width=\textwidth]{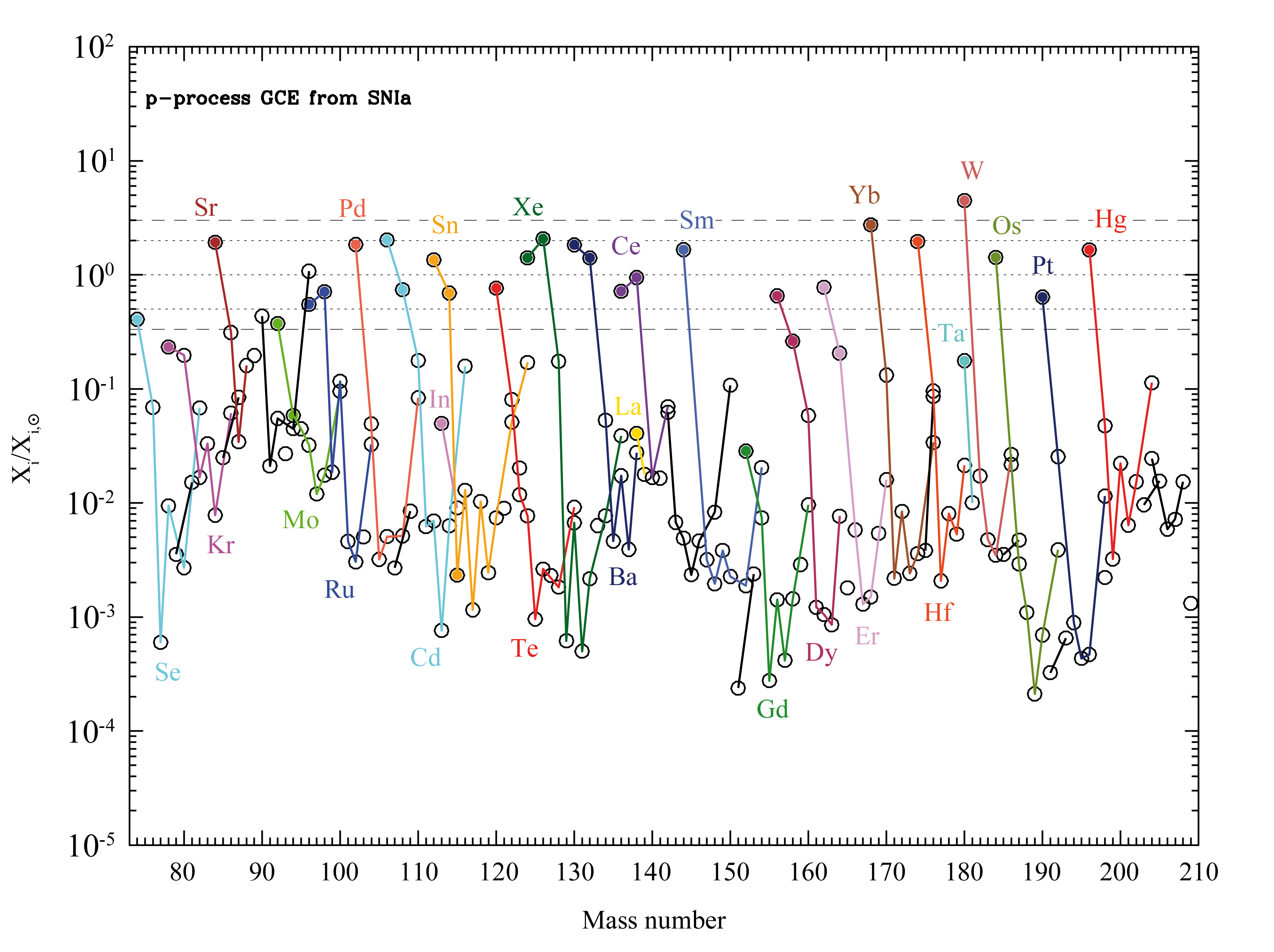}
\caption{Galactic chemical evolution of the $p$~process based on 2D SNe~Ia, taken at the epoch of Solar System formation. Filled dots are for the 35 isotopes classically defined as $p$-only. The isotopes of each element are connected by a line, and for each element a different colour is adopted.}
\label{fig:chemev}
\end{figure}

Recent work by Refs.~\cite{travaglio:11,travaglio:14,travaglio:15} demonstrated that SNe~Ia may be an important source for $p$~nuclei. They used an extended nuclear network with 1024 species from neutron and proton up to $^{209}$Bi combined with neutron-, proton-, and $\alpha$-induced reactions. In particular, Ref.~\cite{travaglio:15} showed that SNe~Ia could be responsible for at least 50\% of all $p$~nuclei required for galactic chemical evolution, under the assupmtions that SNe~Ia are responsible for 2/3 of the solar $^{56}$Fe and the delayed detonation model represents the typical SNIa with a frequency of 70\%~\cite{li:11}. Fig.~\ref{fig:chemev} shows the obtained production factors in comparison to the Solar System composition, using a simple chemical evolution code and accounting for 2D SNe~Ia alone. Most of the $p$-nuclei abundances are reproduced within a factor of three. Only $^{113}$In, $^{115}$Sn, $^{138}$La, $^{152}$Gd and $^{180}$Ta are far below the average production factor of the other $p$~nuclei. As we discussed in the previous sections, they are made by other processes, while the $\gamma$ process is not producing them efficiently. The relative low abundance of the $p$~isotope $^{158}$Dy should be analyzed in the framework of present nuclear uncertainties of the $\gamma$~processes.

Interesting to notice is the still puzzling origin of $^{94}$Mo. Present nuclear uncertainties cannot account for the factor of ten deficiency in the $^{94}$Mo abundance relative to other $\gamma$-process abundances. The $^{94}$Mo production was found to depend on the seeds in the Mo isotopes as well as on the density at which the photodisintegration process occurs. This leaves room for possible variations in the hydrodynamic history of the contributing explosive tracers, which could change the relative abundance distribution. 

Ref.~\cite{travaglio:14} also discuss the production of the short-lived radionuclides $^{92}$Nb, $^{146}$Sm and $^{97,98}$Tc by single degenerate SNIa stars. Using a simple GCE code, they show that a significant fraction of the $p$ extinct radionuclides $^{92}$Nb, $^{146}$Sm, and $^{96,98}$Tc found in meteorites could have been produced by the $\gamma$-process in SNe~Ia. A detailed investigation of nuclear uncertainties affecting the reaction rates producing and destroying $^{92}$Nb, $^{92}$Mo, and $^{146}$Sm found that the calculated $^{146}$Sm/$^{144}$Sm ratio was compatible with the meteoritic value if the $^{148}$Gd($\gamma$,$\alpha$) rate is either based on a fit to the ($\alpha$,$\gamma$) cross sections~\cite{somorjai:98} or on the recent rate including an additional reaction channel~\cite{rauscher:13}. Concerning $^{92}$Nb, the most important reactions affecting the $^{92}$Nb/$^{92}$Mo ratio were discussed and the impact of their nuclear uncertainties were explored. The $^{92}$Nb/$^{92}$Mo ratio predicted by GCE at 9.2 Gyr ranges between \mbox{$1.66\times10^{-5}$} and $3.12\times10^{-5}$ due to the nuclear uncertainties. This demonstrates that the meteoritic value can be reproduced within these uncertainties. The authors conclude that SNe~Ia can play a key role in explaining meteoritic abundances of the extinct radioactivities $^{92}$Nb and $^{146}$Sm, but also that nuclear uncertainties still have considerable impact. However, as discussed in \S \ref{sec:observations} and in Ref.~\cite{lugaro:16}, this scenario is challenged by observations of the radionuclide $^{53}$Mn.  

\begin{figure}
\centerline{\includegraphics[width=0.8\textwidth]{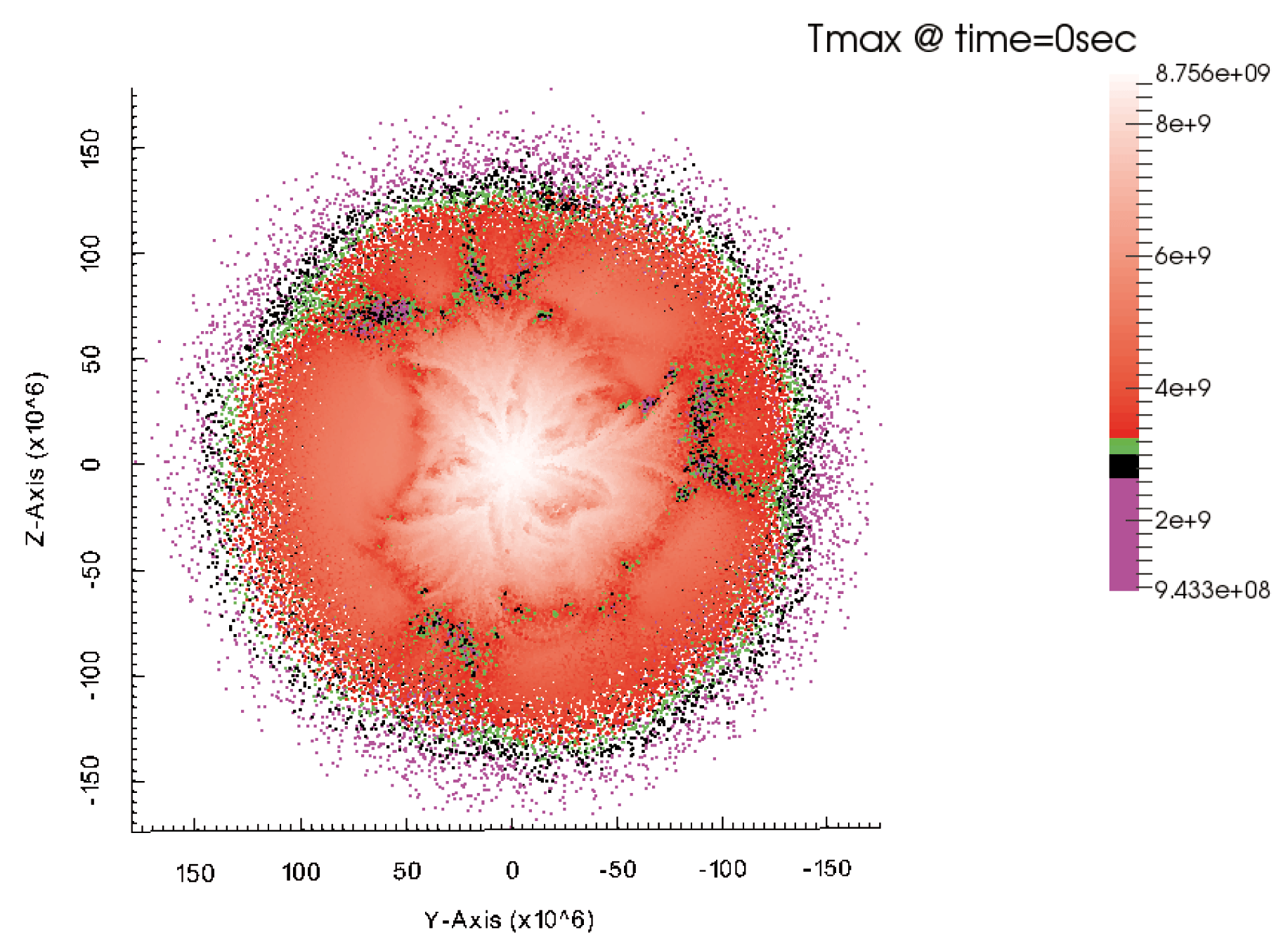}}
\caption{Distribution of the 1 million tracers in the 3D model {\sl N100}~\cite{ropke:12,seitenzahl:13} at the time when the explosion started. The tracers are colored by the maximum temperature reached by each particle during the explosion.}
\label{fig:3Dmodel}
\end{figure}

\subsection{The production of the $p$ nuclei in 2D and 3D}

Recent calculations~\cite{seitenzahl:13} follow the explosion phase of a Chandrasekhar-mass WD not only by means of 2-dimensional but also 3-dimensional hydrodynamic simulations. Fig.~\ref{fig:3Dmodel} shows a 3D initial distribution of tracer particles for the model {\sl N100}s in Refs.~\cite{ropke:12,seitenzahl:13}. The tracers undergoing $\gamma$-process nucleosynthesis are color-coded in pink, black, and green according to the maximum temperature they reach during the explosion. Comparing Fig.~\ref{fig:2Dmodelsnapshots} and Fig.~\ref{fig:3Dmodel}, one can see that the morphology of the two models is different. The 3D model has multiple ``fingers'' extending into the star, while the 2D model has a layered, nearly spherical structure and a well defined zone ahead of the deflagration flame. The deflagration ashes are clearly visible as the hottest material, and the detonation products are the smoothly varying uniform temperature patches. It seems that the ``fingers'' are connected to the deflagration and that most of the material from these ``fingers'' is situated up near the tops of the large deflagration plumes. Thus in 3D, because of the existence/persistence of small localized structures, the deflagration continues to burn to lower density along certain directions before being engulfed by the detonation. This causes a different morphology in 3D as compared to 2D. It should also be noted that due to the asymmetry the structures are actually ``rings'' and not ``blobs''. In conclusion, as far as the $\gamma$-process conditions are concerned, the 3D model has a more complex structure, which cannot be recovered in 2D.

\begin{figure}
\centerline{\includegraphics[width=\textwidth]{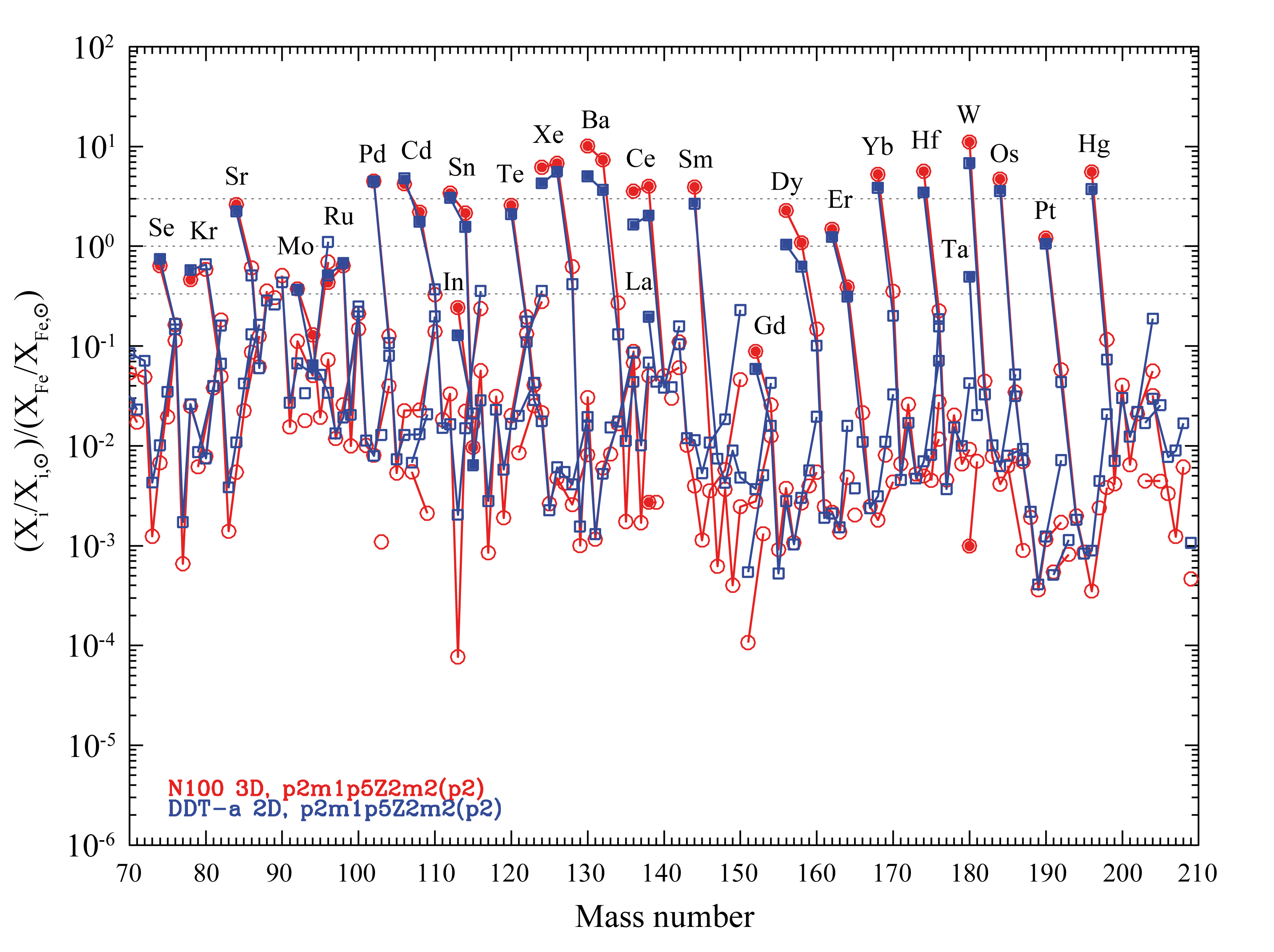}}
\caption{$p$-process yields normalized to solar values and to Fe obtained for solar metallicity SNIa models: deflagration-to-detonation transition model DDT-a (blue squares, compare Fig.~\ref{fig:2Dmodelsnapshots}) and {\sl N100}~\cite{ropke:12,seitenzahl:13} (red circles). Filled squares and circles are for $p$-only nuclei.}
\label{fig:3d_2d}
\end{figure}

The resulting abundances for 2D and 3D are shown in Fig.~\ref{fig:3d_2d}, normalized to Fe and to the solar values~\cite{lodders:09}. The blue filled squares and the red filled circles are the $p$~nuclei from $^{74}$Se up to $^{196}$Hg, for 2D and 3D, respectively. As one can see, the resulting nucleosynthesis is very similar in 2D and in 3D, with a few exceptions. The biggest difference shows up in $^{180}$Ta and $^{138}$La, where the abundances are about a factor of 10 higher in 2D. But these isotopes are synthesized at the lowest temperature of all $p$~nuclei, i.e. in the outermost regions of the star, where the 3D model is not very well resolved by tracers. Smaller differences of a factor between two and three, but with the opposite trend (abundances are higher in 3D than in 2D), are seen for $^{94}$Mo, $^{113}$In,$^{115}$Sn, $^{130,132}$Ba, $^{136,138}$Ce, $^{152}$Gd, and
$^{156,158}$Dy. All these isotopes are mostly produced at temperatures between 2.6$\times 10^9$ and 2.8$\times 10^9$~K.

\subsection{Final remarks for the $\gamma$ process in SNe Ia}

Type Ia supernova may be an important source for the production of $p$~nuclei in the Universe. At present, the most uncertain parameter for the $\gamma$-process production of $p$~nuclei is the $s$-process seed distribution obtained during the WD accretion stage before reaching the Chandrasekhar mass. Consistent stellar calculations have not been published so far. The results by Ref.~\cite{travaglio:15} and presented in this section, e.g., were obtained using $s$-process seeds from the He-intershell in the AGB evolutionary phase. These abundances can differ quite significantly from the abundances build during the WD accretion phase. Efforts are well underway to solve these limitations.

\section{Nuclear physics for the $\gamma$ process}
\label{nuclear: rene}

The reaction network to describe complete p-process scenarios includes a huge number of reaction rates. The experimental database is scarce: (i) The amount of available sample material is usually limited, in particular for radioactive samples. (ii) The cross sections are small in the astrophysically relevant energy range (Gamow window) due to the Coulomb barrier between the charged reaction partners. High-intensity particle beams are needed at experimental facilities for the signature of the reactions to exceed the background.

Reaction rates are generally calculated in the framework of the Hauser-Feshbach statistical model~\cite{RaT00,arnould:03}. In the last decade, different experimental studies have been carried out to test the reliability of these predictions and their sensitivity to the nuclear physics input parameters, such as $\gamma$-ray strength functions and particle-nucleus optical potentials (OP). Radiative capture reactions were observed with in-beam and activation methods to understand the influence of different particle-nucleus OPs and to learn about the inverse reactions being involved in $\gamma$-process nucleosynthesis. This approach was followed for protons~\cite{SLD08,GOY09,KGE07}, $\alpha$-particles~\cite{somorjai:98,SBL07,YGO09} and neutrons~\cite{DDH10,MDD10} at various facilities.

Sensitivity studies were performed to determine the rates with the largest effect on the final abundances. The impact of theoretical reaction rates and their uncertainties on the $\gamma$-process nucleosynthesis has been analyzed for CCSNe with general uncertainty studies~\cite{rapp:06,rauscher:06,rauscher:15}, or with sensitivity studies of the $\gamma$-process production focused on specific mass regions (e.g., for $^{92}$Mo and $^{94}$Mo isotopes~\cite{gobel:15}).

Experimental progress usually results immediately in more robust $\gamma$-process yields. An example is the analysis of the $^{146}$Sm/$^{144}$Sm ratio in the early solar system~\cite{rauscher:13a,travaglio:14}, where $^{144}$Sm is a stable $p$~nuclei and $^{146}$Sm is a radioactive long-lived product of the $\gamma$ process (see also discussion in \S \ref{sec:observations} and \ref{pprocess-snia}). In Ref.~\cite{travaglio:14}, a local uncertainty study has been performed for the production of $^{92}$Nb, $^{92}$Mo, $^{144}$Sm and $^{146}$Sm in SNIa explosions. General sensitivity studies for the $\gamma$ process in SNe Ia are not yet available.

\subsection{$\gamma$-induced reactions}

Reactions induced by $\gamma$-rays are the dominating mechanism to produce proton-rich nuclei during the $\gamma$~process. The corresponding reaction cross sections and stellar rates are therefore needed for all isotopes along the photodisintegration paths. While charged-particle-induced reactions probably play a role in the production of the lighter $p$~nuclei, the $p$~nuclei beyond mass 100 are almost exclusively produced via $\gamma$-induced reactions and subsequent $\beta^{+}$/EC reaction chains.

If the stellar temperatures are high enough (T$\ge$1.5$\times$10$^{9}$ K, see the next sections) to allow ($\gamma$,p),($\gamma$,n), and ($\gamma,\alpha$) reactions, also low-lying states in the nuclei are populated. The population of these states follows the Boltzmann distribution: 

\begin{equation}\label{boltzmann}
  \frac{N_{i}}{N_{\mathrm{GS}}} = \frac{g_{i}}{g_{\mathrm{GS}}} \mbox{e}^{-\frac{E_{i}}{kT} }
\end{equation}

where $E_{i}$ is the energy of the excited state, while the energy of the ground state is zero. The factors $g_{i}=2J_{i}+1$ account for the degeneracy of the energy level.

Even though the population of these states is typically much smaller than that of the ground state (degeneracy ignored), the likelihood of $\gamma$-induced particle-emission reactions on these states is increased. The nucleus is already closer to the particle emission energy $Q$, hence, the necessary additional energy to dissociate a particle is much smaller. More photons with lower energies are available since the high-energy tail of the Planck-distribution decreases exponentially with the photon energy: 

\begin{eqnarray}
  N_\gamma (E)     & \propto & \frac{E^{3}}{\mbox{e}^{E/kT}-1} \\
          & \approx & {\mbox{e}^{-\frac{E}{kT}}}
\end{eqnarray}

Each state has the same contribution to the stellar reaction rate - if all other selection rules can be neglected - independent of its position between ground state and the particle emission energy:

\begin{equation}\label{rates}
  \frac{R_{i}}{R_{\mathrm{GS}}} \approx \frac{N_{i}}{N_{\mathrm{GS}}} \frac{N_{\gamma}(Q-E_{i})}{N_{\gamma}(Q)} \approx 1
\end{equation}

Despite their importance, reactions on excited states are extremely difficult to determine experimentally, since these states are typically very short-lived. Terrestrial experiments can only determine ground state rates or, in exceptional cases, rates on long-lived states (isomers).

\subsubsection{Experimental techniques}

If the product of the investigated ($\gamma$,p),($\gamma$,n) or ($\gamma,\alpha$) reaction is unstable, the activation technique in combination with real photons can be applied. A sample of typically 1~g is irradiated with $\gamma$-rays. The produced radioactivity~\cite{SMV03} or the nuclei themselves~\cite{DFK10} can be detected afterwards. Electrons hitting a copper target emit bremsstrahlung and can be used as a $\gamma$-ray source. The $\gamma$-energy can be derived from the difference of the beam energy and the electron energy if the energy of the decelerated electron is measured. This method is called photon tagging and can in principle be used to determine the reaction cross sections as a function of $\gamma$-energy~\cite{SLG10}.
Alternatively, photons can be produced by Compton scattering: High-energy electrons interact with laser light and transfer energy via inverse Compton effect~\cite{SSF14}.

The described methods are only applicable to stable isotopes since the available $\gamma$-ray fluxes at current facilities are too low to investigate small samples. One option is to use radioactive ion beams. The interaction of ions moving at relativistic speed with the Coulomb field of a heavy nucleus at rest can be interpreted as the interaction with a virtual photon field~\cite{BaR96,BeG10}. This process is called Coulomb dissociation (CD) if light particles are emitted. The experimentally determined CD cross section can be converted into ($\gamma$,p),($\gamma$,n), or ($\gamma,\alpha$) cross sections by applying the virtual photon theory~\cite{LLA14}. In addition, the principle of detailed balance can be applied, which allows to contrain the time-inversed capture reactions. This method is successful in particular for low nuclear level densities. (Only few levels are available in the compound nucleus as for light nuclei or nuclei close to nuclear shell closures.)~\cite{STH06,RHF08}

In general, it is much more reliable to derive the astrophysically interesting stellar ($\gamma$,p),($\gamma$,n),($\gamma,\alpha$)-rates from terrestrial (p,$\gamma$),(n,$\gamma$),($\alpha,\gamma$) cross sections than from terrestrial $\gamma$-induced cross sections~\cite{Rau11}. The impact of reactions on excited states, which are not accessible in the laboratory, can be estimated if the exothermal direction is measured.

\subsection{(p,$\gamma$) reactions}

Proton capture reactions on heavy isotopes are only rarely important for the $\gamma$~process. Only a few isotopes have a significant destruction via (p,$\gamma$) reactions and even fewer actually reach a (p,$\gamma$)-($\gamma$,p) equilibrium. However, the knowledge of the (p,$\gamma$) cross sections is valuable since it allows to determine (p,$\gamma$)-reaction rates under stellar conditions. Applying the principle a detailed balance, the (p,$\gamma$)-reaction rate can be used to determine the important stellar ($\gamma$,p)-reaction rates.

\subsubsection{Experimental techniques}\label{sec:pg_exp}

Typical proton capture measurements are performed close to or at the upper end of the Gamov window~\cite{DHK06} using the in-beam method~\cite{LFH87,SEN12} or, if possible, the activation technique~\cite{SaK97,BSK98}. Due to the current limitations in accelerator and detection methods, both techniques are limited to isotopes with half lives of at least ten years.

It is possible to measure proton-induced reactions at low energies at the Frankfurt Neutron Source at the Stern Gerlach Zentrum (FRANZ) facility at the Goethe University Frankfurt, Germany~\cite{RCH09,WCD10}. FRANZ, which is currently under construction, is based on a high-intensity proton accelerator. Neutrons in the energy range up to 500~keV are produced via the $^{7}$Li($p,n$) reaction. However, one beam line is foreseen to use the proton beam directly for proton-induced reactions. In the current layout, the FRANZ facility provides protons in an energy range of 1.8 MeV to 2.2 MeV. Thus, it covers the low-energy part of the Gamow window at typical temperatures for $\gamma$-process nucleosynthesis. For the measurement of proton-induced reactions the accelerator will be operated in cw mode using the RFQ and IH structure. The proton beam current in cw mode reaches up to 20~mA in the current design of FRANZ. Compared to Van-de-Graaff accelerators the current is enhanced by a factor of 100 to 1000. Therefore, measurements of very small reaction cross sections or the usage of small samples of radioactive material is possible. Experiments to investigate (p,$\gamma$) reactions in the region around the neutron shell closure at $N=50$ are planned, which includes the production chain of $^{92}$Mo via proton captures.  

New experiments to investigate direct (p,$\gamma$) reactions are currently under development at the experimental storage ring (ESR) at GSI Helmholtzzentrum f\"ur Schwer\-ionen\-forschung in Darmstadt, Germany. The reactions are measured in inverse kinematics at energies close to the Gamow window of the $\gamma$~process~\cite{MAB15}. Ions are stored in the ring and interact with a hydrogen gas target. In particular the successful development of microjet gas targets for the ESR allows significantly higher target densities than previously achievable~\cite{KPW09}. If successful, this method can be applied to isotopes with half lives down to minutes. Also, it can potentially be applied at other facilities, e.g. the ion storage ring at REX-ISOLDE~\cite{GLR12}.

\subsection{($\alpha,\gamma$) reactions}

Under stellar conditions, $\alpha$-capture cross sections are even smaller than proton capture cross sections. Therefore, $\alpha$-capture reactions do not occur on heavy elements in stars. However, similar to the proton captures, ($\alpha,\gamma$) cross sections are very important to determine the stellar ($\gamma,\alpha$) rates~\cite{GSG14}. So far, all of the relevant $\alpha$-induced cross sections on heavy elements have been measured using the activation technique, see for example Ref.~\cite{KSG11} and~\cite{KSR14}. Also here, the usage of microdroplet targets~\cite{KPW09} in combination with ion storage rings might open a new era of experiments~\cite{LBB13}.

\subsection{(n,$\gamma$) reactions}

In the $\gamma$~process, the neutron evaporation process stalls at neutron-deficient isotopes leading to a (n,$\gamma$)-($\gamma$,n) equilibrium~\cite{ArG03}. Neutron capture cross sections for neutron-deficient isotopes are of immediate importance for the understanding of the $\gamma$~process. However, the determination of (n,$\gamma$) cross sections is also important to determine stellar ($\gamma$,n) rates. Experiments are challenging since most of the corresponding isotopes are unstable, in particular at higher temperatures and elements beyond $Z=50$.

\subsubsection{Experimental techniques}\label{sec:ng_exp}

Neutron capture cross section can be determined by the time-of-flight (TOF) method~\cite{RLK14} or the activation technique~\cite{RAH03,RHF08}. 
The TOF method usually requires rather big and isotopically pure samples, which sometimes prohibits a successful TOF-experiment. 

The FRANZ~\cite{RCH09,WCD10,RHH04} facility (see section~\ref{sec:pg_exp}) will investigate neutron-induced reactions via the TOF and the activation technique. It will host the strongest neutron source for astrophysical research world-wide. However, the TOF method reaches an important limit when going to shorter and shorter half lives. Depending on the decay properties, the emitted particles from the sample limit the sample size. Hence, isotopes below a certain facility-specific half life cannot be investigated. 

The measurement of the neutron capture cross section in inverse kinematics would enable the study of nuclei with half lives down to minutes. The isotope under investigation could be accelerated and interact with neutrons in a reactor core~\cite{ReL14}. This method would work for center-of-mass energies down to $\approx$~100~keV. While this covers only the high-energy regime of the $s$ process ($kT$=10-90~keV), it would cover the interesting energy range of the $p$ process almost completely ($kT$=200-400~keV).

\subsection{$\beta$ decays}

Stellar $\beta$-decay rates can differ from the terrestrial rates by orders of magnitude~\cite{takahashi:87}. The rate uncertainties can be of particular relevance in the pre-SN component of the $\gamma$ process (see Ref.~\cite{rauscher:02}, and next section). The underlying physics mechanisms involve excited nuclear states and changes of the phase space of the parent or daughter systems.

Electron capture can occur on excited states which are energetically not allowed on earth~\cite{LaM98}, and $\beta$-decays occur from thermally excited states. These decay rates cannot be measured in the laboratory. Both effects may alter the decay rates by a few orders of magnitude~\cite{takahashi:87,AAA04}. For the theoretical calculations of stellar rates, Gamow-Teller strength distributions $B$(GT) for low lying states are needed~\cite{FFN80,langanke:00,LaM03}. 

Charge-exchange reactions, like the (p,n) reaction, allow access to these transitions and can serve as input for rate calculations. In particular, there exists a proportionality between (p,n) cross sections at low momentum transfer (close to 0$^{\circ}$) and $B$(GT) values,

\begin{equation}
\frac{d\sigma^{\mathrm{CE}}}{d\Omega}(q=0)=\hat{\sigma}_{\mathrm{GT}}(q=0)B(\mathrm{GT}),
\end{equation}

where $\hat{\sigma}_{\mathrm{GT}}(q=0)$ is the unit cross section for GT transitions at q=0~\cite{TGC87}. Experiments have to be carried out in inverse kinematics with radioactive ion beams in order to access GT distributions for unstable nuclei. This requires the detection of low-energy neutrons at large angles relative to the incoming beam~\cite{SPZ12,LAC11}. This method is in principle also possible for other charge-exchange reactions~\cite{NZA14}.

The second important mechanism altering the observed decay rates results from the ions being almost completely ionized in the stellar plasma. This obviously alters the electron capture rates, but even more so the $\beta^{-}$-rates if the Q-value is small. The effects of ionization can only be investigated experimentally if the ions are stored over long time using ion storage rings. Such measurements are difficult to perform, but extremely valuable. Only very few cases have been measured so far~\cite{BFF96}. 

There has been no attempt to measure the nuclear capture of a free electron so far. In principle, such a measurement could be performed using an ion storage ring where the ions are stored at rather high energies of several hundred AMeV. After a quick change of the electron-cooler energy, the electron cooler could serve as the electron target for the stored ions. It would be necessary to disentangle the products of nuclear electron captures from atomic electron captures (de-ionization). This is possible using particle detectors with ion-identification capabilities, like \mbox{$\Delta E$-$E$}-detectors.

Except for few special cases, like $^{26}$Al or $^{79}$Se, where the half lives of the long-lived first excited states have been experimentally determined, stellar $\beta$-decay rates and electron captures have to rely on theoretical model calculations~\cite{coc:00,klay:88,takahashi:87,langanke:00,pruet:03,aikawa:05}. This will be the case for quite some time also for $\gamma$-process nucleosynthesis simulations.

\section{Summary and final discussion}
\label{sec: conclusions}

We have reviewed the production of $p$~nuclei by the $\gamma$~process in stars. Different $p$-process scenarios likely contributed to the production of all the $p$~nuclei observed in the Solar System, but the $\gamma$~process is the only known process that can produce proton-rich isotopes beyond the Pd mass region. 

Obervations from meteoritic data form the basis for nucleosynthesis simulations to understand the origin of the $p$~nuclei. We summarized how isotopic abundance anomalies point at anomalies in the $s$~seed distributions for the $\gamma$~process, or at contributions from other $p$-process scenarios to the observed $p$~abundances. 

We discussed the $\gamma$~process in core-collapse supernovae CCSN and its dependence on the seed distribution, which is determined by the rates of fusion reactions and the weak $s$~process during the evolution of the progenitor, as well as the metallicity and the mass of the progenitor. CCSNe fail to reproduce the solar $p$~nuclei abundances, especially the light $p$~nuclei \isotope[92,94]{Mo} and \isotope[96,98]{Ru} are highly underproduced. However, we highlight that the use of \isotope[16]{O} as a reference isotope is problematic because of the difficulty to define its solar abundance. Also, more recent stellar models show a large variation of the $p$-nuclei yields, affected by the details of the stellar structure evolution of massive stars before the SN explosion, and by the uncertainties of CCSN models. The variations may reduce the underproduction of the $p$~nuclei. 

Thermonuclear supernovae SNe~Ia represent good astrophysical candidates for the production of the $p$~nuclei. Recently, different works have investigated $p$-process nucleosynthesis in the framework of two-dimensional single-degenerate scenarios. The main uncertainties lie in the $s$-process seed distribution obtained before the thermonuclear explosion of the white dwarf. The $\gamma$ process in 3D SNIa models needs to be fully explored, as well as the impact of different strengths of the deflagration phase during the explosion. Furthermore, the production of $p$~nuclei should be analyzed in detail also for alternative SNIa progenitors, like sub-Chandrasekhar models and WD mergers (double-degenerate scenarios). 

The contribution of the $\gamma$~process in core-collapse supernovae and in thermonuclear supernovae to the solar $p$~abundances is yet unclear. Both scenarios may provide comparable contributions to the solar abundances, or one may dominate over the other. The contributions have to be determined by galactic chemical evolution simulations using stellar abundance yields from both CCSNe and SNe~Ia. This has been done already for SNe~Ia, but not yet for CCSNe.

Most of the reactions during the $\gamma$~process involve unstable nuclei. The nuclei are either in excited states, which are always unstable, or the ground state itself is unstable. It is therefore extremely challenging to measure the stellar rates in laboratories. More and more facilities around the world are capable of producing radioactive beams of different qualities and intensities. In combination with new detection techniques, a grid of experimentally determined rates will probably be available within the next decade. A major challenge, however, remain the $\gamma$-induced rates because of the importance of excited states. The strategy for the foreseeable future must therefore be to improve the nuclear models capable of predicting the corresponding stellar rates.

After more than fifty years of research, the production of the $p$~nuclei still carries several mysteries and open questions that need to be answered: How many $p$~processes contributed to the solar abundances? What is the relative contribution from CCSNe and SNe Ia? What are the dominant sources in the Mo-Ru region? The answers to these question might be in reach within the next years, when improved nuclear data, data for exotic nuclei, the latest generation of CCSN and SNIa simulations, as well as GCE studies enable more robust predictions for the $\gamma$~process.

\newpage 

\section*{Acknowledgements}

NuGrid acknowledges significant support from NSF grants PHY 02-16783 and PHY 09-22648 (Joint Institute for Nuclear Astrophysics, JINA), NSF grant PHY-1430152 (JINA Center for the Evolution of the Elements) and EU MIRG-CT-2006-046520. The continued work on codes and in disseminating data is made possible through funding from STFC and EU-FP7-ERC-2012-St Grant 306901 (RH, UK), NSERC Discovery grant (FH, Canada), and an Ambizione grant of the SNSF (MP, Switzerland). MP acknowledges support from the ``Lendulet-2014'' Programme of the Hungarian Academy of Sciences and from SNF (Switzerland).
\mbox{NuGrid} data is served by Canfar/CADC. 
The numerical calculations by C.T. have been supported by the B$^2$FH Association. 
R.R. and K.G. acknowledge significant support from the Helmholtz International Center for Fair (HIC for FAIR), DFG (SO907/2-1) and BMBF (05P15RFFN1). This research has received funding from the European Research Council under the European Unions's Seventh Framework Programme (FP/2007-2013) / ERC Grant Agreement n. 615126.

\newpage 

\bibliographystyle{unsrt}
\bibliography{p-review}

\end{document}